\documentclass[a4paper,11pt]{article}
\pdfoutput=1
\usepackage{jcappub}
\usepackage{subfig}
\usepackage{caption}
\usepackage{hyperref}
\usepackage{amsmath}
\usepackage{mathtools}
\usepackage{upgreek}
\usepackage{tablefootnote}
\usepackage{soul}
\usepackage{color}
\usepackage{verbatim}
\usepackage{float}
\usepackage{bm}
\usepackage[usenames,dvipsnames,svgnames,table]{xcolor}

\definecolor{darkred}{RGB}{125, 5, 25}

\definecolor{darkgreen}{RGB}{0,100,0}

\title{From Primordial Black Holes Abundance to Primordial Curvature Power Spectrum (and back)} 

\author[a,b,c]{Alba Kalaja,}
\emailAdd{a.kalaja@rug.nl}

\author[b,d]{Nicola Bellomo,}
\emailAdd{nicola.bellomo@icc.ub.edu}

\author[a,e,f]{Nicola Bartolo,}
\emailAdd{nicola.bartolo@pd.infn.it}

\author[a,e]{Daniele Bertacca,}
\emailAdd{daniele.bertacca@pd.infn.it}

\author[a,e,f,g]{Sabino Matarrese,}
\emailAdd{sabino.matarrese@pd.infn.it}

\author[b]{Ilia Musco,}
\emailAdd{iliamusco@icc.ub.edu}

\author[b,h]{Alvise Raccanelli,}
\emailAdd{alvise.raccanelli@cern.ch}

\author[b,i]{Licia Verde}
\emailAdd{liciaverde@icc.ub.edu}

\affiliation[a]{Dipartimento di Fisica e Astronomia G. Galilei, Universit\`a degli Studi di Padova, I-35131 Padova, Italy.}
\affiliation[b]{ICC, University of Barcelona, IEEC-UB, Mart\'i  i Franqu\`es, 1, E-08028 Barcelona, Spain.}
\affiliation[c]{Van Swinderen Institute for Particle Physics and Gravity, University of Groningen, Nijenborgh 4, 9747 AG Groningen, The Netherlands.}
\affiliation[d]{Dept. de  F\'isica Qu\`antica i Astrof\'isica, Universitat de Barcelona, Mart\'i  i Franqu\`es 1, E-08028 Barcelona, Spain.}
\affiliation[e]{INFN, Sezione di Padova, via F. Marzolo 8, I-35131, I-35131 Padova, Italy.}
\affiliation[f]{INAF - Osservatorio Astronomico di Padova, vicolo dell'Osservatorio 5, I-35122 Padova, Italy.}
\affiliation[g]{Gran Sasso Science Institute, viale F. Crispi 7, I-67100 L'Aquila, Italy.}
\affiliation[h]{Theoretical Physics Department, CERN, 1 Esplanade des Particules, CH-1211 Geneva 23, Switzerland.}
\affiliation[i]{ICREA, Pg. Lluis Companys 23, Barcelona, E-08010, Spain.} 

\abstract{
In the model where Primordial Black Holes (PBHs) form from large primordial curvature (C) perturbations, i.e., CPBHs, constraints on PBH abundance provide in principle constraints on the primordial curvature power spectrum. This connection however depends necessarily on the details of PBH formation mechanism. In this paper we provide, for the first time, constraints on the primordial curvature power spectrum  from the latest limits on PBH abundance, taking into account all the steps from gravitational collapse in real space to PBH formation. In particular, we use results from numerical relativity simulations and peak theory to study the conditions for PBH formation for a range of perturbation shapes, including non-linearities, perturbation profile and a careful treatment of smoothing and filtering scales. We then obtain updated PBH formation conditions and translate that into primordial spectrum constraints for a wide range of shapes and abundances. These updated constraints cover a range of scales not probed by other cosmological observables. Our results show that the correct and accurate modelling of non-linearities, filtering and typical perturbation profile, is crucial for deriving meaningful cosmological implications.
}

\begin{document}

\begin{flushright}
CERN-TH-2019-056
\end{flushright}
\vfil

\maketitle

\section{Introduction}
\label{sec:introduction}
During the first two runs of the LIGO-Virgo observatory, a considerable fraction of detected events~\cite{abbott:firstligodetection, abbott:secondligodetection, abbott:thirdligodetection, abbott:fourthligodetection, abbott:fifthligodetection, abbott:O1O2allevents, venumadhav:O1O2newevents} shows two characteristics that were generally unexpected by part of the community: large progenitors masses ($\gtrsim 20\ M_{\odot}$) and low binary effective spin. 

Even if such massive progenitors were (by some) expected to be the first detected sources~\cite{belczynski:massivebhsmergers, dominik:massivebhsmergersI, dominik:massivebhsmergersII, dominik:massivebhsmergersIII} and are not incompatible with classical stellar/binary evolution~\cite{belczynski:highmassbhI, belczynski:highmassbhII, belczynski:highmassbhIII}, this fact suggested that detected black holes could also have an origin different from the standard end-point of stellar evolution and that they may constitute a significant fraction of the dark matter~\cite{bird:pbhasdarkmatter, clesse:pbhmerging, sasaki:pbhasdarkmatter}. Moreover, black holes of primordial origin, PBHs hereafter, are expected to have low spins, as recently showed in Refs.~\cite{takeshi:spinpbhs, mirbabayi:spinpbhs, deluca:spinpbhs}, hence they would produce binaries with values of the effective spin parameter compatibles with the observed ones. The observed merger rate is on the upper end of the predicted range for stellar progenitors~\cite{fragione:globularclusters} (even though there are still many uncertainties), and at least some contribution from primordial objects would reconcile theory with observations. Furthermore, PBHs might constitute the seeds of the super-massive black holes~\cite{volonteri:smbhformation, latif:smbhformation} that inhabit the center of galaxies~\cite{fan:smbhdetection, mortlock:smbhdetection, wu:smbhdetection, banados:smbhdetection}.

Given the interest in this potential dark matter candidate and the rich phenomenology of black holes, a large number of observational constraints on the abundance of PBHs have been obtained so far. They cover a remarkable portion of the allowed mass range and they include constraints coming from gravitational lensing effects~\cite{katz:femtolensingconstraint, griest:keplerconstraint, niikura:microlensingconstraint, tisserand:microlensingconstraint, calchinovati:microlensingconstraint, alcock:microlensingconstraint, mediavilla:microlensingconstraint, wilkinson:millilensingconstraint, zumalacarregui:supernovaconstraint}, dynamical effects~\cite{carr:evaporationconstraint, raccanelli:evaporationconstraint, ballesteros:evaporationconstraint, poulter:evaporationconstraint, graham:whitedwarfconstraint, quinn:widebinaryconstraint, brandt:ufdgconstraint, nakamura:pbhmergerrate, raidal:pbhmergerrate, alihaimoud:pbhmergerrate, magee:mergerrate, raidal:pbhmergerrateII}, accretion effects~\cite{ricotti:cmbconstraint, gaggero:accretionconstraints, alihaimoud:pbhaccretion, poulin:cmbconstraint, bernal:cmbconstraint, nakama:mudistorsionsconstraints, hektor:accretionconstraint, manshanden:accretionconstraint, hutsi:accretionconstraint} and effects on large-scale structure~\cite{afshordi:lssconstraint, murgia:lssconstraint}. Two mass ranges remain still open, around~$10^{-15}\ M_\odot$ and~$10^{-12}\ M_\odot$. Nonetheless, these constraints are not conclusive because they are computed for monochromatic mass distributions and they involve a variety of assumptions, see e.g., Refs.~\cite{bellomo:emdconstraints, carr:extendedmassdistributionI, carr:extendedmassdistributionII}. Therefore, the model in which PBHs constitute at least a non negligible fraction of the dark matter is still allowed by observations.

The idea that an overdense region of the primordial Universe could collapse gravitationally to form a black hole was proposed already fifty years ago~\cite{zeldovich:pbhsformation, hawking:pbhsformation, carr:pbhsformation, chapline:pbhformation}. Several mechanisms to produce such overdensities have been suggested, including cosmic topological defects~\cite{hawking:pbhsfromstrings, garriga:pbhsfromstrings, caldwell:pbhsfromstrings, matsuda:pbhsfromnecklaces, lake:pbhsfromnecklaces, caldwell:pbhsfromwalls, hawking:pbhsfrombubbles, crawford:pbhsfrombubbles, moss:pbhsfrombubbles}, (interacting) dark matter clumps~\cite{shandera:pbhfromdarkmatter} or large curvature perturbations generated during inflation~\cite{ivanov:pbhsfrominflation, bellido:pbhsfrominflation, ivanov:pbhsfrominflationII}. In the latter formation mechanism, curvature perturbations are generated during inflation, hence they carry a substantial amount of information about the Early Universe. In particular, there are a plethora of inflationary scenarios able to generate PBHs in the late Universe, see e.g.,~Refs.~\cite{leach:runningmassmodel, drees:runningmassmodel, drees:runningmassmodelII, kawasaki:axioncurvatonmodel, kohri:axioncurvatonmodel, bellido:inflectionpointmodel, germani:inflectionpointmodel, kannike:doubleinflationmodel, motohashi:slowrollbreaking, ballesteros:loopcorrectionsmodel, ozsoy:stringtheorymodel, cicoli:stringtheorymodel, dalianis:alphaattractorsmodel, bhaumik:inflectionpointmodel}. In this paper we concentrate on this scenario of PBHs generated by primordial curvature (C) perturbations, CPBHs.

CPBHs, apart from providing a dark matter candidate and being the seeds of super-massive black holes, can provide insights on the first moments of our Universe. It has already been established that at cosmological scales ($k\lesssim\mathcal{O}(1)\ \mathrm{Mpc}^{-1}$) the primordial curvature power spectrum is almost scale-invariant, both in model-dependent and model-independent analyses, see e.g., Ref.~\cite{ravenni:powerspectrumconstraints}. Moreover, during the past two decades, the amplitude and the tilt of the primordial curvature power spectrum has been measured with high accuracy~\cite{spergel:wmapcosmoparams, ade:planckcosmoparams2013, ade:planckinflation2013}. On the other hand we still have very little information about the primordial curvature power spectrum on small scales ($k>\mathcal{O}(1)\ \mathrm{Mpc}^{-1}$). Several authors have proposed different ways to probe such scales, including CMB spectral distortions~\cite{chluba:powerspectrumconstraints}, analyses of Silk damping effects~\cite{jeong:powerspectrumconstraints}, exploiting WIMP properties~\cite{josan:ucmhconstraints, bringmann:powerspectrumconstraints} (in the last case assuming they are the main component of dark matter), reconstructing quasar light curves~\cite{karami:powerspectrumconstraints} or through the detection of gravitational waves generated by large scalar perturbations, see e.g., Refs.~\cite{bellido:secondordergws, cai:secondordergwsI, bartolo:secondordergws, bartolo:secondordergwsII, unal:secondordergws, byrnes:steepestgrowth, inomata:secondordergws, yuan:secondordergws, cai:secondordergwsII} and references therein for constraints coming from ongoing (PTA~\cite{lentati:epta, arzoumanian:nanograv, shannon:ppta}) and future (SKA~\cite{moore:gwsensitivitycurves, janssen:gwsfromska}, LISA~\cite{moore:gwsensitivitycurves, amaroseoane:gwsfromlisa}) experiments.

An additional method to set constraints on the amplitude and shape of the power spectrum consists in using PBH abundance~\cite{carr:pbhmassspectrum, carr:powerspectrumconstraintsI, carr:powerspectrumconstraintsII}. CPBH formation requires at least mildly non-linear fluctuations to form during radiation domination, hence it requires an inflationary dynamics that deviates significantly from the standard slow-roll paradigm. It is generally accepted that in the simplest standard, single-field slow-roll inflationary models, initial perturbations are very close to Gaussian and their power spectrum is an almost scale invariant power law; hence perturbations large enough to go non-linear in the early Universe are exceedingly rare. Therefore constraints on PBH abundance can be translated into constraints on the Early Universe physics. This connection was pioneered in Ref.~\cite{josan:powerspectrumconstraints} and later extended in Refs.~\cite{cole:powerspectrumconstraints, carr:earlymatterdomination, dalianis:powerspectrumconstraints} to include an early matter-dominated era. We refer the interested reader also to Refs.~\cite{mifsud:powerspectrumconstraints, satopolito:powerspectrumconstraints}, where the authors report constraints on the primordial power spectrum amplitude from the most updated PBH abundance constraints, and to Ref.~\cite{akrami:powerspectrumuncertaintities}, where the authors investigated the effects of critical collapse and non-sphericities on the primordial power spectrum constraints.

In the previous literature a series of approximations and shortcuts were used. Given the potential implications of an accurate and robust connection between PBH abundance constraints and early Universe physics, we argue that these approximations should be revisited. The goal of this work is to extend previous works and to put constraints on the primordial curvature power spectrum on a firmer theoretical ground. We aim to do this by building on and improving upon previous analyses as outlined below in section~\ref{sec:executive_summary}.


\section{Executive Summary}
\label{sec:executive_summary}
This work is based on three pillars: the numerical simulations needed to assess the conditions under which PBHs form, the cosmological connection, fundamental to link the properties of the individual peak eventually forming a CPBH to the statistics of random fields and in particular their correlation functions, and peak theory, used to assess CPBH abundance and its link to primordial statistical properties. Each one of these pillars has a dedicated section,~\ref{sec:pbh_formation}, \ref{sec:cosmology_connection} and~\ref{sec:peaks_theory} for numerical simulations, cosmological connection and peak theory, respectively. These elements constitute the fundamental building blocks we use to reconstruct the primordial power spectrum.

We improve upon previous analyses by reducing the number of approximations (also in light of recent theoretical developments) and providing new insights on each of the building blocks as follows. We go beyond the linear approximation for the curvature perturbation in modelling CPBH formation and improve the reliability of the estimate of the critical threshold the perturbation has to overcome to collapse. We include information about the profile of the initial density perturbation. We clarify the role of smoothing scales and their relation to the underlying physics and we propose a filtering recipe that respects all the relevant physics. Finally, we go beyond Press-Schechter theory to connect PBH abundance to the primordial power spectrum and adopt the most recent PBH abundance constraints which have changed significantly since the time of Ref.~\cite{josan:powerspectrumconstraints}.

In section~\ref{sec:pbh_formation} we treat the details of CPBH formation. We perform an advanced study of the effects of non-linearities. In particular we study the impact of the linear approximation of curvature on the typical scale of the collapsing region, on the overdensity profile, on the mass of the final object and on the criterion used to assess whether a PBH forms or not. We prove that none of these quantities is accurately computed using linear theory. 

In section~\ref{sec:cosmology_connection} we further develop cosmological perturbation theory to be applicable in the context of CPBH formation and to include non-linear effects which are non-negligible. We motivate, on a physical basis, how the filtering/smoothing procedure should be done to avoid introducing artificial features on the filtered field. While numerical simulations treat one density perturbation at the time, cosmological perturbation theory treats the entire density field (made by the superposition of many density perturbations) at once. The density field statistical properties are determined by the primordial curvature power spectrum (and possibly higher-order statistics) and by non-linearities. Here we provide a fully analytical method to include non-linearities and  primordial non-Gaussianity contributions to the density field statistics.

In section~\ref{sec:peaks_theory} we connect the results found using numerical simulations to the statistical properties of the density field, which ultimately determines the abundance of PBHs. We comment on how the statistical properties of the density field should be evaluated during radiation-domination and under which conditions the shape of the peak in the density field is connected to statistical properties of the field itself.

We conclude in section~\ref{sec:primordial_power_spectrum_reconstruction}, where we provide the most updated and accurate limits on the primordial curvature power spectrum amplitude allowed if CPBHs constitute the maximum fraction of the dark matter consistent with observations. Furthermore, we show to which extent the initial conditions of numerical simulations, corresponding to the threshold for CPBH formation, can be used to reconstruct the shape of the primordial power spectrum. 

In this work we use natural units $c=\hbar=G=1$ unless otherwise specified.


\section{Primordial black holes formation}
\label{sec:pbh_formation}
The gravitational collapse of density perturbations in the radiation-dominated era and the subsequent formation of PBHs are highly non-linear processes. Hence their study requires numerical simulations~\cite{nadezhin:pbhsimulation, bicknell:pbhsimulation, novikov:pbhsimulation}. After these pioneering works, the collapse of initial perturbations in the form of primordial curvature fluctuations, was numerically studied in Refs.~\cite{niemeyer:pbhmassII, shibata:compactionfunction}, followed some years afterwards by an extensive numerical analysis by one of us using an explicit Lagrangian hydrodynamics code developed and used in Refs.~\cite{musco:pbhformation,polnarev:curvatureprofiles, musco:criticalcollapse, musco:selfsimilarity, musco:pbhthreshold} and, recently, by other authors in Refs.~\cite{escriva:numericalsimulationsI, escriva:numericalsimulationsII}. More details about the code and the result of these simulations are discussed in appendix~\ref{app:numerical_simulations}. Here we rely on results of this code and these simulations.
 
Numerical simulations of PBH formation have always assumed spherical symmetry. This assumption is quite natural in this context because large perturbations, as in the case of those generating CPBHs, are expected to be quasi-spherical (see also section~\ref{sec:peaks_theory}); therefore we will continue assuming spherical symmetry\footnote{Small deviation from non-spherical perturbations could play an important role when computing PBH abundance, however in this work we follow the standard approach.}. The simplest form of the metric in a spherically symmetric spacetime is
\begin{equation}
ds^2 = -A^2(t,r)dt^2 + B^2(t,r)dr^2 + R^2(t,r)d\Omega^2,
\label{eq:spherical_symmetric_metric}
\end{equation}
where $t$ is the cosmic time, $r$ is a comoving radial coordinate, $A$, $B$ and $R$ are strictly positive functions, $d\Omega$ is the solid angle measure. The function $R$ is also called areal radius and it measures the physical distance of a point of the space-time with coordinates $(t,r)$ from the centre of symmetry.


\subsection{Curvature and density perturbations in the super-horizon regime}
\label{subsec:super_horizon_regime}
Formation of a CPBH requires a cosmological perturbation large enough to collapse, forming an apparent horizon~\cite{helou:apparenthorizon, faraoni:foliation} which is obtained from initial conditions characterized by non-linear curvature perturbations. To use the standard description of cosmological adiabatic perturbations behaving as pure growing modes, these initial conditions must be set on super-horizon scales, where the length scale of the perturbation(s) we are considering must be much larger than the cosmological horizon at initial time. This is easy to envision if initial curvature perturbations are generated in the context of cosmological inflation.

In this regime the curvature perturbations are conserved (time-independent) because pressure gradients are negligible and an analytic treatment, usually called the gradient expansion or long-wave length approximation\footnote{In this description the exact solution is expanded in a power series of a small parameter that is conveniently identified with the ratio between the Hubble radius and the length-scale of the perturbation.}, is possible~\cite{salopek:longwavelength, shibata:compactionfunction}. The metric in equation~\eqref{eq:spherical_symmetric_metric} can then be written using a spherically-symmetric spatial curvature perturbation~$K(\tilde{r})$~\cite{polnarev:curvatureprofiles}
\begin{equation}
ds^2 \simeq -dt^2 + a^2(t)\left[ \frac{d\tilde{r}^2}{1-K(\tilde{r})\tilde{r}^2} + \tilde{r}^2d\Omega^2 \right],
\label{eq:K_metric}
\end{equation}
where $a(t)$ is the scale factor and $\tilde{r}$ is a comoving radial coordinate. This expression is approximated because here for simplicity we neglect the time-dependent components of the metric perturbations, which are small on super-horizon scales. However these components are taken into account when the initial condition of numerical simulations are specified. 

For a more expert reader we point out that, although this approach reproduces the time evolution of linear perturbation theory on super-horizon scale, it also allows one to consider non-linear curvature perturbations if the spacetime is sufficiently homogeneous and isotropic on large scales~\cite{lyth:zetacurvature}. This is equivalent to say that pressure gradients are negligible and shows that the large initial curvature perturbations, as required for PBH formation, has to appear already at zero order in the background form of the metric.

In a cosmological framework it is more convenient to use a different parametrisation of the curvature perturbation, for instance by using the curvature perturbation on comoving hypersurfaces~$\mathcal{R}$ or the curvature perturbation on uniform energy density hypersurfaces~$\zeta$.\footnote{In cosmology there are different notation conventions regarding the curvature perturbation. Throughout this work we follow the one of Refs.~\cite{lidsey:curvatureconvention, lyth:curvatureconvention, malik:curvatureconvention}; however there are alternative conventions, for instance the one used by the WMAP and Planck Collaborations, see e.g., Refs.~\cite{komatsu:cosmologicalinterpretation, ade:planckng2013}.} Both can be interpreted as perturbations of the scale factor~$a(t)$. Here we choose to work with the latter, where equation~\eqref{eq:K_metric} becomes~\cite{lyth:zetacurvature}
\begin{equation}
ds^2 \simeq -dt^2 + a^2(t) e^{-2\zeta(\hat{r})} \left[ d\hat{r}^2 + \hat{r}^2d\Omega^2 \right],
\label{eq:cosmo_metric}
\end{equation}
valid in this form during radiation-domination only on super-horizon scales. Different parame-trisations of the curvature perturbation yield different parametrisation of the radial comoving coordinate (see Ref.~\cite{harada:differentgauges} for an exhaustive discussion on different metrics in the context of CPBH formation), and comparing the two forms of the metric above one finds they are related by~\cite{musco:pbhthreshold}
\begin{equation}
\begin{aligned}
\tilde{r} &= \hat{r}e^{-\zeta(\hat{r})}, \\
K(\tilde{r})\tilde{r}^2 &= \hat{r} \frac{d\zeta(\hat{r})}{d\hat{r}} \left[ 2 - \hat{r}\frac{d\zeta(\hat{r})}{d\hat{r}} \right].
\end{aligned}
\end{equation}

\begin{figure}[t]
\centerline{
\includegraphics[width=1.0\columnwidth]{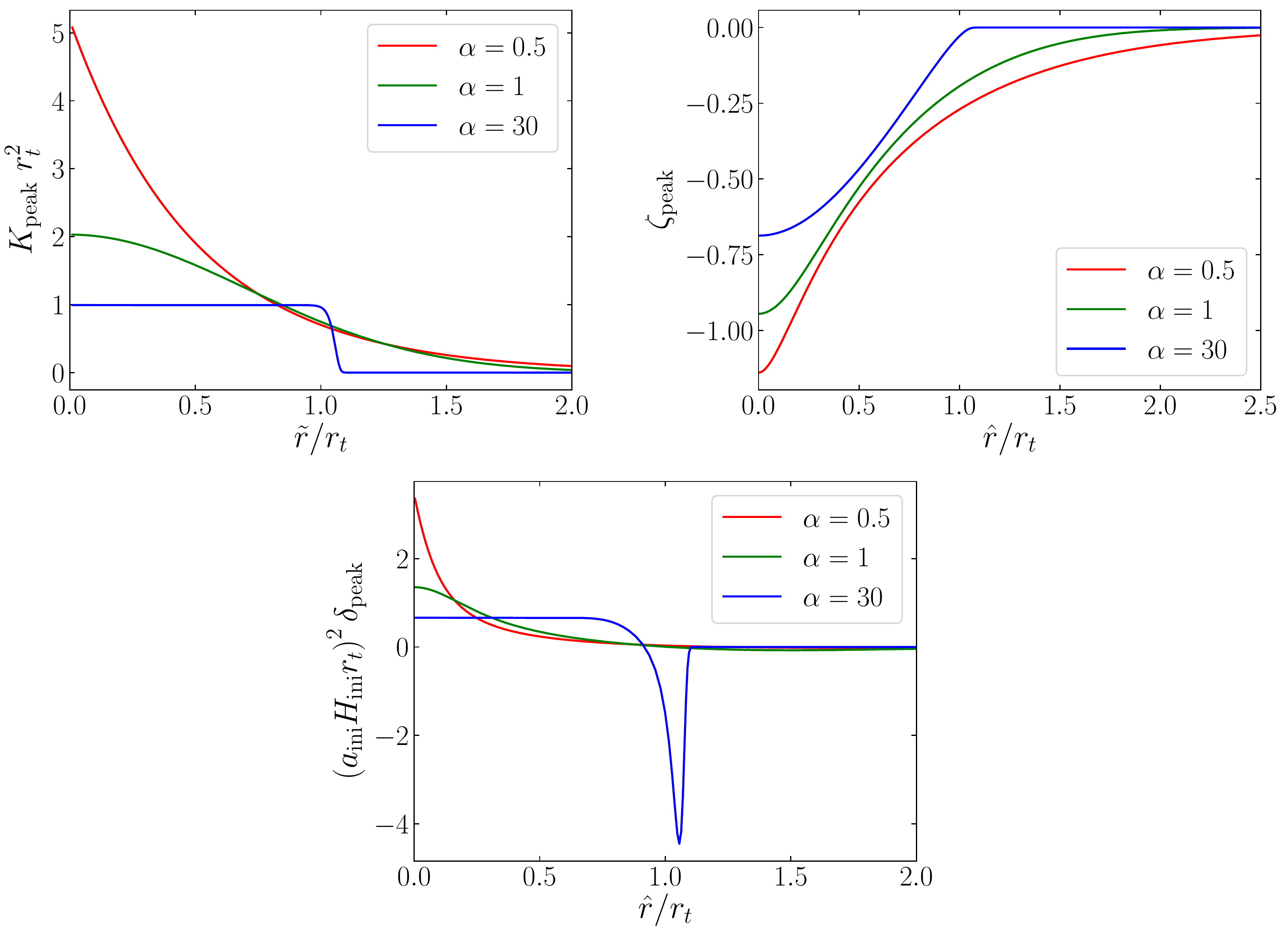}}
\caption{\textit{Upper left panel:} curvature profile~$K_\mathrm{peak}r^2_t$ at initial conditions, on super-horizon scales, for different values of the shape parameter~$\alpha$. We use $\mathcal{A}_\mathrm{peak}r_t^2=5.18,\ 2.03,\ 0.99$ for $\alpha=0.5,\ 1.0,\ 30$, respectively. While the peak amplitude~$\mathcal{A}_\mathrm{peak}$ changes when the typical scale~$r_t$ varies, the quantity $\mathcal{A}_\mathrm{peak}r_t^2$ is constant. Initial conditions are set up super-horizon, so that the typical scale of the perturbation~$r_t$ is much larger than the coming horizon at initial time, i.e.,~$a_\mathrm{ini}H_\mathrm{ini}r_t\gg 1$. \textit{Upper right panel:} Curvature profiles~$\zeta_\mathrm{peak}$, corresponding to the $K$-curvature peaks of the upper left panel, at initial time~$t_\mathrm{ini}$ for different values of the parameter~$\alpha$. Also in this case the typical scale of the perturbation is~$r_t$. \textit{Lower panel:} Corresponding overdensity profiles at initial time~$t_\mathrm{ini}$.}
\label{fig:curvature_profiles}
\end{figure}

From now onwards we assume that the spatial metric perturbations in equations~\eqref{eq:K_metric} and~\eqref{eq:cosmo_metric} describe a peak centred in the coordinates origin. In Refs.~\cite{polnarev:curvatureprofiles, musco:pbhthreshold} the authors analysed the gravitational collapse of density perturbations generated by peaks of the $K$-curvature perturbation, whose profile is parametrised as
\begin{equation}
K_\mathrm{peak}(\tilde{r}) = \mathcal{A}_\mathrm{peak} \exp\left[-\frac{1}{\alpha}\left(\frac{\tilde{r}}{r_t}\right)^{2\alpha}\right],
\label{eq:K_profile}
\end{equation}
where $\mathcal{A}_\mathrm{peak}$ is the $K$-curvature peak amplitude, $\alpha$ describes the steepness of the peak profile and~$r_t$, as it will become clearer later, sets the typical scale of the peak. We choose this one-parameter family of profiles because it allows us to study both steep ($\alpha\to 0$) and flat peaks ($\alpha\to\infty$), as we show in the upper left panel of figure~\ref{fig:curvature_profiles}. This family of profiles assumes a spatially flat background at infinity, i.e., $K(\tilde{r})\to0$ when $\tilde{r}\to\infty$. According to the super-horizon regime prescription, the scale~$r_t$ has to be much larger than the comoving cosmological horizon~$r_\mathrm{hor}=(aH)^{-1}$ at initial time~$t_\mathrm{ini}$, i.e.,~$a_\mathrm{ini} H_\mathrm{ini} r_t\gg 1$, where~$H$ is the Hubble expansion rate, $a_\mathrm{ini}=a(t_\mathrm{ini})$ and~$H_\mathrm{ini}=H(t_\mathrm{ini})$.
The simulations show that, if the perturbation amplitude, controlled by the parameter~$\mathcal{A}_\mathrm{peak}$, is larger than a suitable threshold, these curvature profiles lead to CPBH formation. That is, if certain conditions on the  peak profile  are satisfied, these curvature profiles generate overdensities that, after crossing the cosmological horizon, are large enough to overcome pressure forces, collapse and form PBHs. These conditions will be discussed in \S~\ref{subsec:formation_criterion}.

On super-horizon scales a constant shift in~$\zeta$ is equivalent to a rescaling of the radial coordinate; therefore, without any loss of generality, we can take~$\zeta(\hat{r})\to 0$ when~$\hat{r}\to\infty$. Hence for the family of profiles under consideration, a peak in the $K$-curvature becomes a trough in the $\zeta$-curvature perturbation which reads as
\begin{equation}
\zeta_\mathrm{peak}\left[\hat{r}(\tilde{r})\right] = \int_{\infty}^{\tilde{r}} \frac{dr}{r} \left[\frac{1}{\sqrt{1-K_\mathrm{peak}(r)r^2}} - 1\right],
\label{eq:zeta_profile}
\end{equation}
reported in the upper right panel of figure~\ref{fig:curvature_profiles}. Despite the na\"ive idea that peaks in the curvature field correspond to peaks in overdensity, in this case peaks in the metric of equation~\eqref{eq:K_metric} correspond to troughs in the metric of equation~\eqref{eq:cosmo_metric}, even if both perturbations generate a peak in the energy density\footnote{If we use the curvature perturbation on comoving hypersurfaces~$\mathcal{R}\simeq-\zeta$ instead of the curvature perturbation on uniform energy density hypersurfaces~$\zeta$, then peaks in the $K$-curvature correspond to peaks in the $\mathcal{R}$-curvature and both generate peaks in the density.}~$\rho$. The~$\zeta$-curvature perturbation generates an overdensity perturbation~$\delta_\mathrm{peak}=\rho/\bar{\rho}-1$, where~$\bar{\rho}$ is the background energy density, given by 
\begin{equation}
\begin{aligned}
\delta_\mathrm{peak}(t,\hat{r}) &= \frac{3(1+w)}{5+3w} \left(\frac{1}{aH}\right)^2 \left(-\frac{4}{3}\right) e^{5\zeta_\mathrm{peak}(\hat{r})/2} \nabla^2 e^{-\zeta_\mathrm{peak}(\hat{r})/2}	\\
&= \frac{4}{9} \left(\frac{1}{aH}\right)^2 e^{2\zeta_\mathrm{peak}(\hat{r})} \left[\nabla^2 \zeta_\mathrm{peak}(\hat{r}) - \frac{1}{2} \nabla \zeta_\mathrm{peak}(\hat{r}) \cdot \nabla \zeta_\mathrm{peak}(\hat{r}) \right],
\end{aligned}
\label{eq:overdensity_cosmo_metric}
\end{equation}
where~$w:=p/\rho$ is the equation of state for a perfect fluid and $p$ is the pressure\footnote{We refer the interested reader to appendix B of Ref.~\cite{helou:apparenthorizon} for more details about the equation of state of a perfect fluid.} (\mbox{$w=1/3$} in radiation-dominated era). As  it can be seen in the lower panel of figure~\ref{fig:curvature_profiles} (and later in figure~\ref{fig:overdensity_profiles}), the peak shows an overdensity in the central region surrounded by an underdense region. We call zero-crossing distance~$\hat{r}_0$ the distance from the peak where $\delta_\mathrm{peak}(\hat{r}_0)=0$, i.e., where the overdensity becomes an underdensity. 

Notice that the relation between the overdensity and curvature perturbations is intrinsically non-linear. A linear relation is recovered only when the curvature perturbation and the gradient of the curvature are small ($|\zeta_\mathrm{peak}|\ll1$, $|\nabla\zeta_\mathrm{peak}|\ll 1$) and we explicitly show in \S~\ref{subsec:validity_linear_approximation} that this approximation is not accurate in the context of CPBH formation. In the following subsection we work with the full non-linear relation between overdensity and curvature.


\subsection{Primordial black hole formation criterion}
\label{subsec:formation_criterion}
Many criteria have been proposed to assess whether a CPBH forms (see Ref.~\cite{musco:pbhthreshold} and references therein for a broad discussion on different criteria). In this work we use the criterion proposed in Ref.~\cite{shibata:compactionfunction}, based on the so called compaction function. This approach allows us to consistently compare curvature and overdensity profiles with different shapes. The compaction function~$\mathcal{C}$ quantifies the magnitude of the gravitational potential and, following Ref.~\cite{musco:pbhthreshold}, we define it as twice the ratio between the mass excess~$\delta M$ inside a sphere of areal radius~$R$ at time~$t$ and the areal radius itself: 
\begin{equation} 
\mathcal{C}(t,r) := 2\frac{\delta M(t,r)}{R(t,r)}.
\end{equation}
Here with  $r$ we refer to a generic comoving variable as in equation~\eqref{eq:spherical_symmetric_metric}, without specifying the particular parametrisation of the curvature profile used.

In the cases where the overdensity perturbation has a single peak, as in those described by equation~\eqref{eq:K_profile}, the compaction function has a maximum at some comoving scale~$r_m$. Moreover, it can be shown that the compaction function is conserved~\cite{musco:pbhthreshold} (i.e., constant in time) on super-horizon scales, where~$R(t,r)H(t)\geq 1$. The condition~$R(t,r)H(t)=1$ defines the horizon crossing time~$t$ of the comoving scale~$r$, hence for any~$t_\star\leq t$ the cosmological horizon is smaller than the scale of interest~$r$ and we have $\mathcal{C}$ constant in time: \mbox{$\mathcal{C}(t_\star,r)=\mathcal{C}(t,r)$.} In the following we take as reference scale~$r_m$ and as reference time~$t_m$, defined implicitly by~$H_m R_m=1$, with~$R_m=R(t_m,r_m)$ and~$H_m=H(t_m)$. In this sense the value of the compaction function can be used as a time-independent measure of the amplitude of the perturbation on super-horizon scales: a CPBH forms if $\mathcal{C}(t_m,r_m)$ is larger than some critical threshold~\cite{shibata:compactionfunction} whose specific value depends on the particular shape of the peak curvature~\cite{musco:pbhthreshold}.

\begin{figure}[t]
\centerline{
\includegraphics[width=1.0\columnwidth]{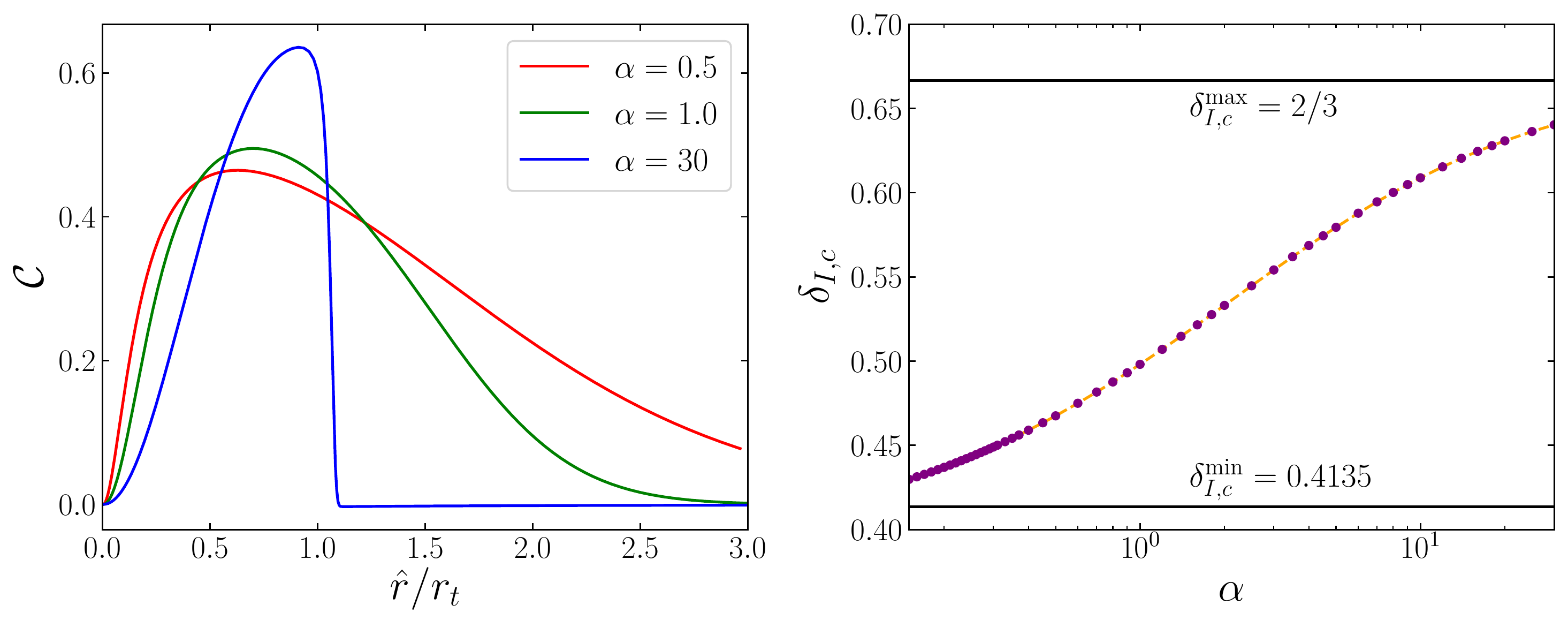}}
\caption{\textit{Left panel:} compaction function~$\mathcal{C}$ corresponding to the curvature and density perturbations of figure~\ref{fig:curvature_profiles}. \textit{Right panel:} Critical threshold as a function of the shape of the perturbation profile. Steep profiles correspond to~$\alpha\to 0$, while flat profiles correspond to~$\alpha\to\infty$. Purple dots represents values found by numerical simulation, interpolated with the orange line.}
\label{fig:critical_collapse}
\end{figure}
 
By comparing equations~\eqref{eq:spherical_symmetric_metric} and~\eqref{eq:cosmo_metric} we find that the areal radius in the second metric reads as~$R = a\hat{r} e^{-\zeta_\mathrm{peak}(\hat{r})}$ and therefore, according to our parametrisation of the peak shape of equation~\eqref{eq:K_profile}, the physical scale of the maximum of the compaction function is $R_m = a_m \hat{r}_m e^{-\zeta_\mathrm{peak}(\hat{r}_m)}$ (with~$a_m=a(t_m)$). This is tightly related to the typical comoving scale of the perturbation $r_t \equiv  \tilde{r}_m = \hat{r}_m e^{-\zeta_\mathrm{peak}(\hat{r}_m)}$ defined by the horizon crossing condition~$a_m H_m r_t = 1$. Qualitatively, we say that at time~$t_m$ the perturbation enters the horizon, i.e., the perturbation enters the horizon when the physical scale~$R_m$ (corresponding to the maximum of the compaction function) is crossing the cosmological horizon. We show in figure~\ref{fig:critical_collapse} the compaction function for the overdensity perturbations appearing in figure~\ref{fig:curvature_profiles}.

For practical purposes, the criterion collapse is often formulated in terms of the integrated overdensity profile~$\delta_I$, given by
\begin{equation}
\delta_I(t,r) = \frac{\delta M(t,r)}{\overline{M}(t,r)} = \frac{3}{R^3(t,r)}\int_0^r dx \frac{dX(t,x)}{dx} X^2(t,x) \delta_\mathrm{peak}(t,x),
\label{eq:definition_deltaI}
\end{equation}
where $X$ and $x$ are physical and comoving coordinates, respectively, and the mass excess 
\begin{equation}
\delta M(t,r) = M(t,r) - \overline{M}(t,r) = 4\pi \int_0^r dx \frac{dX(t,x)}{dx} X^2(t,x) \left[\rho(t,x)-\bar{\rho}(t)\right]
\end{equation}
has been measured with respect to an unperturbed sphere of areal radius~$R$, uniform density~$\bar{\rho}$ (which in the case considered here corresponds to the background cosmological energy density) and enclosed mass~$\overline{M}$. Since we assume spherical symmetry, as in the numerical simulations we use, the mass is defined without ambiguity by the Misner-Sharp mass\footnote{In general one should use the Komar mass~\cite{komar:massingr}, which is equivalent to Kodama mass~\cite{kodama:massingr} and to the Misner-Sharp mass~\cite{misner:massingr} in this context.}~\cite{misner:massingr}.

However the two criteria are not conflicting, in fact, using the definition of compaction function, we have that on super horizon scales $\mathcal{C}(t,r) = H^2(t) R^2(t,r) \delta_I(t,r)$~\cite{harada:differentgauges}. Using the horizon crossing condition~$H_mR_m=1$, our threshold criterion reads as
\begin{equation}
\mathcal{C}(t_m, r_m) = \delta_I(t_m, r_m) > \delta_{I,c}(\alpha),
\end{equation}
where the exact numerical value of the critical threshold~$\delta_{I,c}(\alpha)$ has to be found using numerical simulations (see appendix~\ref{app:numerical_simulations}). We report the critical value of the integrated overdensity amplitude as a function of the parameter~$\alpha$ in the right panel of figure~\ref{fig:critical_collapse}. We note that the critical threshold is shape-dependent and that it takes values between~$\delta^\mathrm{min}_{I,c}=0.4135$ and~$\delta^\mathrm{max}_{I,c}=2/3$ for steep ($\alpha\to 0$) and flat ($\alpha\to\infty$) profiles, respectively. The difference in the value of critical threshold is related to the role of pressure gradients during the non-linear evolution~\cite{musco:pbhthreshold}: a steeper initial profile needs a lower excess of mass to form a PBH because most of the energy density is already located in the centre and the pressure gradients around~$r_t$ are negligible. On the contrary, when the profile is more homogeneous, as for a top-hat profile, the pressure gradients around~$r_t$ are very large and the required value of $\delta_{I,c}$ is higher.

The mass of the resulting CPBH follows the scaling law of critical collapse \cite{niemeyer:pbhmassI} determined by how much the integrated density profile exceeds a critical value and it reads as~\cite{choptuik:pbhmass, niemeyer:pbhmassI, niemeyer:pbhmassII, musco:criticalcollapse}
\begin{equation}
M_\mathrm{PBH} = \mathcal{K}(\alpha) M_\mathrm{hor}(t_m) \left[\delta_{I}(t_m,r_m)-\delta_{I,c}(\alpha)\right]^{\upgamma_\mathrm{crit}},
\label{eq:mass_critical_collapse}
\end{equation}
where~$M_\mathrm{hor}(t_m)=(2H_m)^{-1}$ is the mass contained inside the cosmological horizon at horizon crossing time, $\upgamma_{\rm crit}\simeq 0.36$ is a critical exponent for radiation, which depends only on the equation of state parameter~$w$,\footnote{Notice that the equation of state is not exactly constant during the entire radiation-dominated era. For instance, during the QCD phase transition, the equation of state and the sound speed soften dropping to~$\upgamma\simeq 0.20-0.25$ and the production of PBHs is enhanced~\cite{jedamzik:pbhsqcdI, jedamzik:pbhsqcdII, byrnes:equationofstate, carr:equationofstate}. This change of the equation of state, not any more characterised only by one parameter, generates a deviation from the scaling law, as simulations suggest~\cite{jedamzik:pbhsqcdII}. Since refined simulation of PBH formation during the QCD phase transition investigating are still missing, in this work we consider only the standard case where~$\upgamma\simeq 0.36$.}
~\cite{nielsen:pbhcriticalexponent} while~$\mathcal{K}$ is a numerical coefficient that depends on the specific density profile. This result holds under the condition $M_\mathrm{PBH}\lesssim M_\mathrm{hor}$, i.e., for $\delta_{I}-\delta_{I,c}\lesssim \mathcal{O}(10^{-2})$, beyond these values the scaling law is not very accurate. We discuss further the validity of equation~\eqref{eq:mass_critical_collapse} in \S~\ref{subsec:pbhs_abundance}. It is important to note here that in equation~\eqref{eq:mass_critical_collapse} the estimated values of~$\mathcal{K}$ and~$\upgamma_\mathrm{crit}$ are computed with a~$\delta_I(t_m,r_m)$ which comes from the initial conditions linearly extrapolated and rescaled by the background cosmic evolution (i.e., the effects of pressure gradients are neglected). Also this subtlety will be revisited later, in \S~\ref{subsec:pbhs_abundance}.

As we will see in section~\ref{sec:peaks_theory}, it is useful to re-interpret equation~\eqref{eq:mass_critical_collapse} in terms of the peak amplitude $\delta_{\mathrm{peak},0}=\delta_\mathrm{peak}(t_m,0)$. Also in this case the peak amplitude is linearly extrapolated from initial conditions by using only the cosmic expansion $1/a^2H^2$ factor. The integrated overdensity is related to the overdensity peak amplitude by a shape-dependent, but time-independent, relation $\mathcal{F}_\delta(\alpha)=\delta_{I}(t_m,r_m)/\delta_\mathrm{peak}(t_m,0)$. Hence equation~\eqref{eq:mass_critical_collapse} can be re-written as
\begin{equation}
M_\mathrm{PBH} = \mathcal{K}'(\alpha) M_\mathrm{hor}(t_m) \left[\delta_{\mathrm{peak},0}-\delta_{\mathrm{peak},0,c}(\alpha)\right]^{\upgamma_\mathrm{crit}},
\label{eq:mass_critical_collapse_II}
\end{equation}
where $\mathcal{K}'(\alpha)=\left[\mathcal{F}_\delta(\alpha)\right]^{\upgamma_{\rm crit}}\mathcal{K}(\alpha)$ and the new critical threshold is related to the integrated critical threshold by $\delta_{\mathrm{peak},0,c}(\alpha)=\delta_{I,c}(\alpha)/\mathcal{F}_\delta(\alpha)$.


\subsection{Effects of the linear curvature approximation} 
\label{subsec:validity_linear_approximation} 
It is usually assumed, even in the context of CPBH formation, that both the~$\zeta$-curvature and curvature gradients are small ($|\zeta_\mathrm{peak}|$, $|\nabla\zeta_\mathrm{peak}|\ll 1$) and therefore equations involving them can be linearised. In particular, in equation~\eqref{eq:overdensity_cosmo_metric} it is often assumed that the exponential damping~$e^{2\zeta_\mathrm{peak}}$ and the quadratic gradient correction $(\nabla\zeta_\mathrm{peak}\cdot\nabla\zeta_\mathrm{peak})$ can be neglected, effectively linearising the relation between $\zeta$-curvature perturbation and the overdensity perturbation $\delta$: $\delta^\mathrm{LIN}_\mathrm{peak} \propto \nabla^2\zeta_\mathrm{peak}$.

\begin{figure}[t]
\centerline{
\includegraphics[width=1.0\columnwidth]{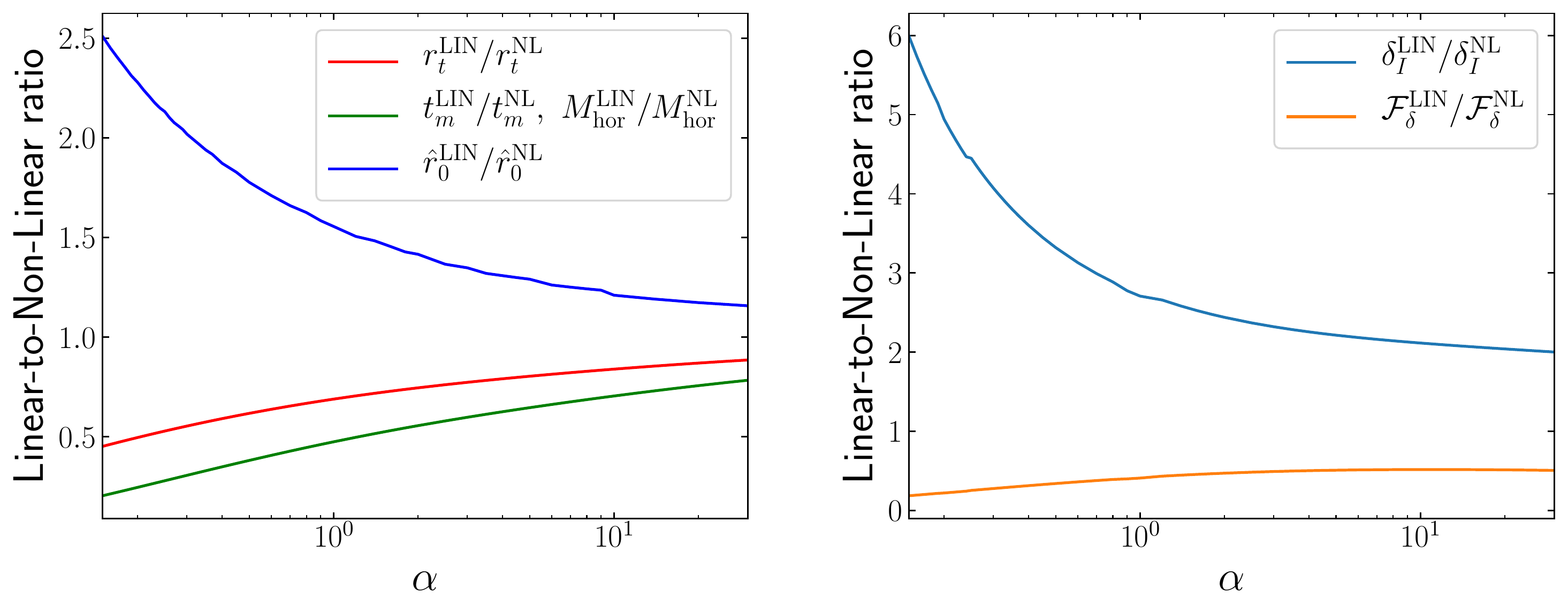}}
\caption{Linear-to-Non-Linear ratio between different physical scales. \textit{Left panel:} ratio of typical comoving scales~$r_t$ of the perturbation, horizon crossing times~$t_m$, masses enclosed inside the horizon~$M_\mathrm{hor}$ and overdensity zero-crossing scales~$\hat{r}_0$. \textit{Right panel:} ratio of integrated density profiles~$\delta_I(t_m,R_m)$ and integrated-density-to-peak-amplitude relation~$\mathcal{F}_\delta$.}
\label{fig:linear_vs_nonlinear}
\end{figure}

Even if at initial time the linear approximation is accurate, at horizon crossing time, when both curvature and overdensity perturbations are of order unity, non-linearities have already produced significant effects. Neglecting non-linearities biases the results obtained at every level of the analysis done in this work (the three pillars of section~\ref{sec:executive_summary}). Hence, we use the full non-linear results coming from numerical simulations.

Since, for simplicity, it is tempting to use the linear approximation -- and it has been used in the literature not infrequently -- in this subsection we explicitly show the effects that the linear approximation generates in the CPBH formation process. To quantify the effects of the linear approximation, we proceed in comparing key quantities evaluated with the full non-linear equation ($\mathrm{NL}$) with the corresponding linear approximation ($\mathrm{LIN}$).

Under linear approximation, the typical scale of the perturbation becomes $r^\mathrm{LIN}_t=\hat{r}_m$, to be compared against the typical non-linear scale $r^\mathrm{NL}_t=\hat{r}_m e^{-\zeta_\mathrm{peak}(\hat{r}_m)}$. Since $\zeta_\mathrm{peak}$ is negative, linearisation underestimates the real size of the perturbation, i.e., $r^\mathrm{LIN}_t/r^\mathrm{NL}_t<1$. Because in radiation-domination the comoving horizon scales as $r_\mathrm{hor}\propto t^{1/2}$ and the horizon crossing condition is $a_m H_m r_t = 1$, linearisation also underestimates the horizon crossing time $t_m$:
\begin{equation}
\frac{t^\mathrm{LIN}_m}{t^\mathrm{NL}_m} = \left(\frac{r^\mathrm{LIN}_t}{r^\mathrm{NL}_t}\right)^2 < 1 .
\end{equation}

\begin{figure}[t]
\centerline{
\includegraphics[width=1.0\columnwidth]{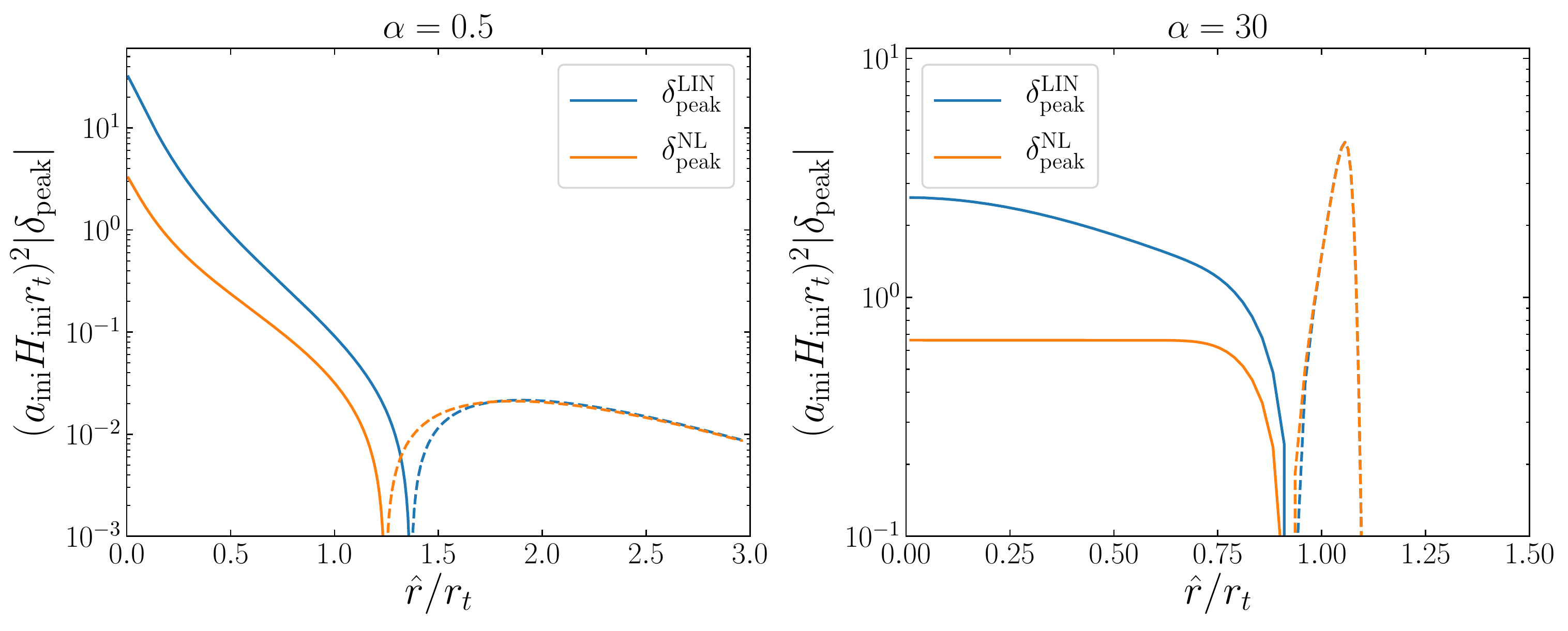}}
\caption{Linear and non-linear overdensity profile of equation~\eqref{eq:overdensity_cosmo_metric} on super-horizon scales, for a steep profiles with $\alpha=0.5$ (\textit{left panel}) and a flat profile $\alpha=30$ (\textit{right panel}). Solid lines indicate overdensity $(\delta_\mathrm{peak}>0)$ while dashed lines indicate underdensities $(\delta_\mathrm{peak}<0)$. The profile is showed in units of the typical scale of the perturbation $r_t=\hat{r}_m$ in the linear approximation or $r_t=\hat{r}_m e^{-\zeta_\mathrm{peak}(\hat{r}_m)}$ in the non-linear case. In both cases the non-linear corrections damps and shrink the overdensity profile significantly with respect to the linear case. We use $\mathcal{A}_\mathrm{peak} r^2_t = 5.18,\ 0.99$ for $\alpha=0.5,\ 30$, respectively.}
\label{fig:overdensity_profiles}
\end{figure}

The inferred mass of the PBH is also affected by linearisation through a different estimate of the horizon crossing time, as can be seen in equation~\eqref{eq:mass_critical_collapse}. The main effect is that the linear approximation underestimates the mass inside the horizon, by a factor
\begin{equation}
\frac{M_\mathrm{hor}^\mathrm{LIN}}{M_\mathrm{hor}^\mathrm{NL}} = \frac{t^\mathrm{LIN}_m}{t^\mathrm{NL}_m} < 1.
\end{equation}
We report in the left panel of figure~\ref{fig:linear_vs_nonlinear} the magnitude of such effects, where we can see that the largest effects are for the steeper profiles. 

The second effect of the linear approximation is to change the non-linear relation between overdensity and $\zeta$-curvature perturbations of equation~\eqref{eq:overdensity_cosmo_metric} yielding: 
\begin{equation}
\delta^\mathrm{LIN}_\mathrm{peak} = \frac{4}{9} \left(\frac{1}{aH}\right)^2\nabla^2 \zeta_\mathrm{peak}.
\label{eq:density_curvature_linear_relation}
\end{equation}
We show in figure~\ref{fig:overdensity_profiles} the comparison between the non-linear overdensity~$\delta_\mathrm{peak}\equiv\delta^\mathrm{NL}_\mathrm{peak}$ and  its linear counterpart~$\delta^\mathrm{LIN}_\mathrm{peak}$ on super-horizon scales. Notice that the profiles are presented in terms of their respective typical scale $r_t$, which we have already showed to be different. Unsurprisingly, the linear approximation clearly fails when curvature and overdensity perturbations are order unity. As we can see from equation~\eqref{eq:overdensity_cosmo_metric}, non-linearities produce two  effects: an exponential damping of the peak given by~$e^{2\zeta_\mathrm{peak}}$ and a change in the tails of the profile given by the gradient correction~$\nabla \zeta_\mathrm{peak} \cdot \nabla \zeta_\mathrm{peak}$. The exponential damping is relevant especially towards the center of the perturbation, where~$|\zeta_\mathrm{peak}|\sim\mathcal{O}(1)$, whereas the gradient-squared correction changes the tails by shifting the zero-crossing scale~$\hat{r}_0$. The change in the absolute value of~$\hat{r}_0$ is showed in the left panel of figure~\ref{fig:linear_vs_nonlinear}. The exponential damping is relevant for the whole range of profile under consideration, while the gradient correction is significant only for steep profiles. Non-linear effects therefore damp and shrink the overdensity profile computed under the linear approximation. At the same time, they also shrink the typical scale, as discussed before. The magnitude of these two competing effects depends on the case under consideration.

Finally, since the linearised overdensity profile is different, the compaction function, the integrated overdensity and the threshold criterion will change as well. We show the linear-to-non-linear ratio of the integrated overdensity profiles~$\delta_I(t_m, r_m)$ and of the integrated-density-to-peak-amplitude relation~$\mathcal{F}_\delta$ in the right panel of figure~\ref{fig:linear_vs_nonlinear}. Since the linearised overdensity is overestimated, it is easier for a perturbation to exceed the critical threshold and to form a CPBH in the linear approximation. Moreover, the mass of the resulting CPBH and, more in general, the CPBHs mass distribution will be shifted with respect to the non-linear case. The magnitude of the shift and the direction (towards lower/higher masses) depend on two competing effects: the change in the horizon crossing time and the change in the threshold criterion. Notice also that the integrated-density-to-peak-amplitude relation~$\mathcal{F}_\delta$ changes as well, hence the numerical coefficient~$\mathcal{K}'(\alpha)$ derived in the linear approximation is smaller.

While the magnitude of all these effects is specific to the family of profiles we have analysed, the qualitative effects remain in general valid, showing that the linear approximation is inadequate to describe the formation of CPBHs. As a consequence, adopting the linear approximation to compute the inferred statistical properties of the overdensity field yield an incorrect estimate of the global PBH abundance. A discussion on these effects can be found in section~\ref{sec:peaks_theory} and in Refs.~\cite{kawasaki:nonlinearities, deluca:nonlinearities, young:nonlinearities}. Since this require a set of preliminary results that are connected to the standard description of cosmological perturbations, we present them after section~\ref{sec:cosmology_connection}.


\section{The cosmology connection}
\label{sec:cosmology_connection}
Our numerical simulations model the collapse of one overdensity peak at a time; however, if CPBHs are the dark matter, we have to construct a self-consistent description of a Universe filled with many of those peaks. Even if the mechanism generating these overdensity perturbations is not known and many of them have been proposed so far, in this work we assume that they are created by large fluctuations of the curvature perturbation, generated during inflation. Hence the information encoded in the overdensity peaks can be connected to the Early Universe physics.

Cosmological perturbation theory provides in principle the necessary framework and tools to connect Early and Late Universe physics. However, in the standard cosmological context and on large cosmological scales, perturbations are typically (almost) linear while we have seen in section~\ref{sec:pbh_formation} that CPBH formation is strongly non-linear and that non-linearities cannot be neglected. Therefore, for this application, standard cosmology perturbation theory has to be extended and modified to account for these effects.

We begin by reviewing cosmological linear theory, which represents our starting point in \S~\ref{subsec:linear_theory}. The first extension we must introduce is connected to the role of the filter functions in the context of CPBH formation, which we discuss in \S~\ref{subsec:filtering_random_fields} and we motivate their use based on physical arguments. The second extension is the inclusion of non-linearities, hence of non-Gaussainities, both primordial (eventually) and due to gravity: their presence is unavoidable precisely because of the non-linear relation between curvature and overdensity and we quantify their effects in \S~\ref{subsec:nongaussianity}.

In this section we always work with the metric defined in equation~\eqref{eq:cosmo_metric}, hence for convenience we drop the ``$\ \widehat{\ }\ $'' symbol on top of spatial comoving coordinates. Notice that in this section we consider generic overdensity and curvature perturbations (and not only peaks, as in section~\ref{sec:pbh_formation}), hence we also drop the ``$\ _\mathrm{peak}\ $'' subscript.


\subsection{Linear theory}
\label{subsec:linear_theory}
In the standard cosmological framework, linear theory can be used and stretched even to study the collapse of massive objects such as halos (see e.g., the Press Schechter approach~\cite{press:pressschechter}). As a warm up exercise, we start by adopting the linear approximation and reviewing the necessary background to be applied to CPBH formation.

In linear theory, as it can be seen in equation~\eqref{eq:density_curvature_linear_relation}, the overdensity perturbation~$\delta$ is uniquely determined by the second derivatives of the $\zeta$-curvature perturbation, \mbox{(i.e., $\delta\propto\nabla^2\zeta$)}. According to the standard interpretation of cosmological perturbation theory, both perturbations are random fields whose properties are determined by the family of $n$-point correlators, e.g., $\left\langle\zeta(\mathbf{x}_1)\cdots\zeta(\mathbf{x}_n)\right\rangle$ or $\left\langle\delta(\mathbf{x}_1)\cdots\delta(\mathbf{x}_n)\right\rangle$, and by the relation between the two random fields. For instance, since the Laplacian is a linear operator, if $\zeta$ is a Gaussian random field then also $\delta$ will be a Gaussian random field.

It is well-known that random fields are neither continuous or differentiable~\cite{rice:correlationprofiles, adler:randomfields, bardeen:peakstheory}, hence it is necessary to smooth out the field on small scales using a filter function, especially to define topological concepts as peaks or troughs of the random field. In full generality, we define a smoothed field, e.g., the overdensity field, as
\begin{equation}
\delta_s(\mathbf{x}) = \int d^3y W_s(|\mathbf{x}-\mathbf{y}|)\delta(\mathbf{y}),
\label{eq:linear_filtering_definition}
\end{equation}
where $W_s$ is a filter function of comoving radius~$s$. The filter function is typically normalized to unity $\left(\int d^3yW_s(|\mathbf{y}|) = 1\right)$ and it can be written in terms of an unnormalized filter function~$w_s(|\mathbf{y}|)$ and a comoving volume normalizing factor $V^\mathrm{com.}_w=\int d^3y w_s(|\mathbf{y}|)$ as $W_s = w_s/V^\mathrm{com.}_w$. 

Since we are working in the linear approximation, where spatial curvature is assumed to be small, the smoothing can be done directly in comoving coordinates. We discuss the smoothing procedure when spatial curvature is not negligible and the appropriate size of the smoothing radius~$s$ in \S~\ref{subsec:filtering_random_fields}. In principle, since the relation between $\zeta$ and $\delta$ is non-linear, smoothing the curvature field is not equivalent to  smoothing directly the over density field. However, at linear order, the two operations are equivalent, in fact by applying the smoothing procedure of equation~\eqref{eq:linear_filtering_definition} to equation~\eqref{eq:density_curvature_linear_relation} we obtain
\begin{equation}
\begin{aligned}
\delta_s(\mathbf{x}) &= \frac{4}{9}\frac{1}{a^2H^2} \int d^3y W_s(|\mathbf{x}-\mathbf{y}|)\nabla^2_{\mathbf{y}}\zeta(\mathbf{y})\\
&= \frac{4}{9}\frac{1}{a^2H^2} \int d^3y \left\lbrace\zeta(\mathbf{y}) \nabla^2_{\mathbf{y}} W_s(|\mathbf{x}-\mathbf{y}|) + \right. \\
&\qquad\qquad\qquad\qquad \left. + \nabla_{\mathbf{y}} \cdot \left[W_s(|\mathbf{x}-\mathbf{y}|)\nabla_{\mathbf{y}} \zeta(\mathbf{y}) -\zeta(\mathbf{y}) \nabla_{\mathbf{y}} W_s(|\mathbf{x}-\mathbf{y}|) \right]\right\rbrace \\
&= \frac{4}{9}\frac{1}{a^2H^2} \nabla^2_{\mathbf{x}}\zeta_s(\mathbf{x}) \, .
\end{aligned}
\label{eq:smoothed_density_curvature_relation}
\end{equation}
The surface contribution vanishes under the fairly general assumption that $W_s$ and its derivative vanish at large scales (as for Top-Hat or Gaussian filter functions), where we use~$\nabla^2_{\mathbf{y}} W_s=\nabla^2_{\mathbf{x}} W_s$ because of the form of the filter function argument and we recognise~$\zeta_s(\mathbf{x}) \equiv \int d^3yW_s(|\mathbf{x}-\mathbf{y}|)\zeta(\mathbf{y})$. Therefore, at linear level, it is completely equivalent to smooth out the overdensity field or the curvature field.

In general, the statistical properties (the $n$-point correlators) of smoothed fields will be different from those of the original unsmoothed field. In particular, the power spectrum of the smoothed field will be that of the unsmoothed field multiplied by the square of the Fourier transform of the kernel. Moreover, the filter function may introduce non-trivial effects in the context of PBHs abundance constraints such as those presented in Ref.~\cite{ando:pbhsmoothingeffects} and later re-analysed in Ref.~\cite{young:windowfunction} (see also \S~\ref{subsec:filtering_random_fields}) or in Ref.~\cite{verde:dmsmoothingeffects} for effects applied to dark matter halos.

Even if so far we have considered the spatial curvature on super-horizon scales as time-independent, on every sub-horizon patch the curvature is actually evolving with time. The evolution of sub-horizon scales is typically described by a transfer function~$\mathcal{T}$. By taking the Fourier transform of equation~\eqref{eq:smoothed_density_curvature_relation} and including pressure effects on sub-horizon scales, we have that in linear theory the overdensity field in Fourier space reads as
\begin{equation}
\widehat{\delta}_s(\tau,\mathbf{k}) = -\frac{4}{9}\frac{k^2}{a^2H^2}\widehat{W}_s(k)\widehat{\mathcal{T}}_\mathrm{LIN}(\tau,k)\widehat{\zeta}(\mathbf{k}),
\label{eq:density_curvature_linear_relation_fourier}
\end{equation}
where~$\tau$ is the conformal time and~$\widehat{W}_s$ and~$\widehat{\mathcal{T}}_\mathrm{LIN}$ are the Fourier transform of the filter and linear transfer functions. Under the linear approximation assumption, in the radiation-dominated era, the transfer function reads as~\cite{dodelson:transferfunction, blais:transferfunction}
\begin{equation}
\widehat{\mathcal{T}}_\mathrm{LIN}(\tau, k) = 3 \frac{\sin(c_sk\tau) - (c_sk\tau)\cos(c_sk\tau)}{(c_sk\tau)^3},
\label{eq:transfer_function}
\end{equation}
where~$c^2_s=1/3$ is the sound speed of the relativistic fluid. As can be seen from equation~\eqref{eq:transfer_function}, pressure effects act as a smoothing and naturally damp perturbations on scales smaller than the sound horizon~$r_s(\tau)=c_s\tau=c_s/(aH)$, i.e., for modes~$k\gg (c_s\tau)^{-1}$.

The statistical properties of the smoothed overdensity field, i.e., the $n$-point functions~$\left\langle\widehat{\delta}_s(\mathbf{k}_1)\cdots\widehat{\delta}_s(\mathbf{k}_n)\right\rangle$ can be computed using equation~\eqref{eq:density_curvature_linear_relation_fourier}, assuming that we know the entire set of $n$-point functions of the curvature field, e.g.,
\begin{equation}
\begin{aligned}
\left\langle \widehat{\zeta}(\mathbf{k}_1)\widehat{\zeta}(\mathbf{k}_2) \right\rangle &= (2\pi)^3\delta^D\left(\mathbf{k}_{12}\right) P_\zeta(\mathbf{k}_1), \\
\left\langle \widehat{\zeta}(\mathbf{k}_1)\widehat{\zeta}(\mathbf{k}_2)\widehat{\zeta}(\mathbf{k}_3) \right\rangle &= (2\pi)^3\delta^D\left(\mathbf{k}_{123}\right) B_\zeta(\mathbf{k}_1, \mathbf{k}_2, \mathbf{k}_3),
\end{aligned}
\end{equation}
etc., where $\delta^D$ is the Dirac delta, $\mathbf{k}_{1\dots n}=\mathbf{k}_{1}+\cdots+\mathbf{k}_{n}$ and $P_\zeta$ and $B_\zeta$ are the curvature power spectrum and bispectrum, respectively. For instance, the two-point function or, equivalently, the power spectrum of the smoothed density field is
\begin{equation}
P_s(\tau, k) = \frac{16}{81}\frac{k^4}{a^4H^4} \widehat{W}^2_s(k) \widehat{\mathcal{T}}^2_\mathrm{LIN}(\tau, k) P_\zeta(k).
\label{eq:smoothed_overdensity_Pk}
\end{equation}


\subsection{Filtering random fields}
\label{subsec:filtering_random_fields}
Filtering is a procedure widely used in signal processing that eliminates the power contained in a range of ``frequencies'' (scales in this case) from some ``signal'' (the overdensity field in the case under consideration). Operationally, this is done by convolving the signal with a filter/smoothing/window function (all the three names have been used in the literature)\footnote{Smoothing is usually the result of a low-pass filter, where high-frequencies are suppressed, but high-pass filters are also useful, where low frequency signals that might mimic an almost-DC mode or long baseline variations are suppressed. Here for example, as it will be clear later,  long wavelength modes on scales much larger than the typical size of the perturbation of interest are considered as DC modes and effectively ignored in the simulations. In certain cases, see \S~\ref{subsec:peak_shape}, these modes have to be cut out.}. A low-pass filter is what is mostly used in cosmological settings and it results in a smoother version of the initial signal. Because of this, in the rest of the section we use interchangeably the term filter and smoothing function.

As already emphasized in Refs.~\cite{rice:correlationprofiles, adler:randomfields, bardeen:peakstheory}, a filtering procedure is absolutely necessary to define concepts such as peaks or troughs of  random fields, which require the  field to be at least differentiable. In this sense, the filter function is just a mathematical artefact we introduce to treat analytically random fields, therefore it is fundamental for this procedure not to bias the statistical properties of the random field (or, in case it is unavoidable, one should asses the magnitude of the bias). Notice that in the Press-Schechter formalism of large-scale structure the filter function is used to define the mass of the object of interest (the dark matter halo). In the CPBHs case this does not apply; in fact the mass of the CPBHs is specified by the shape of the overdensity, the horizon crossing time of the perturbation and by how much the integrated overdensity (or the height of the overdensity peak) exceeds some critical threshold (see equations~\eqref{eq:mass_critical_collapse} and~\eqref{eq:mass_critical_collapse_II}). 

\begin{figure}[t]
\centerline{
\includegraphics[width=1.0\columnwidth]{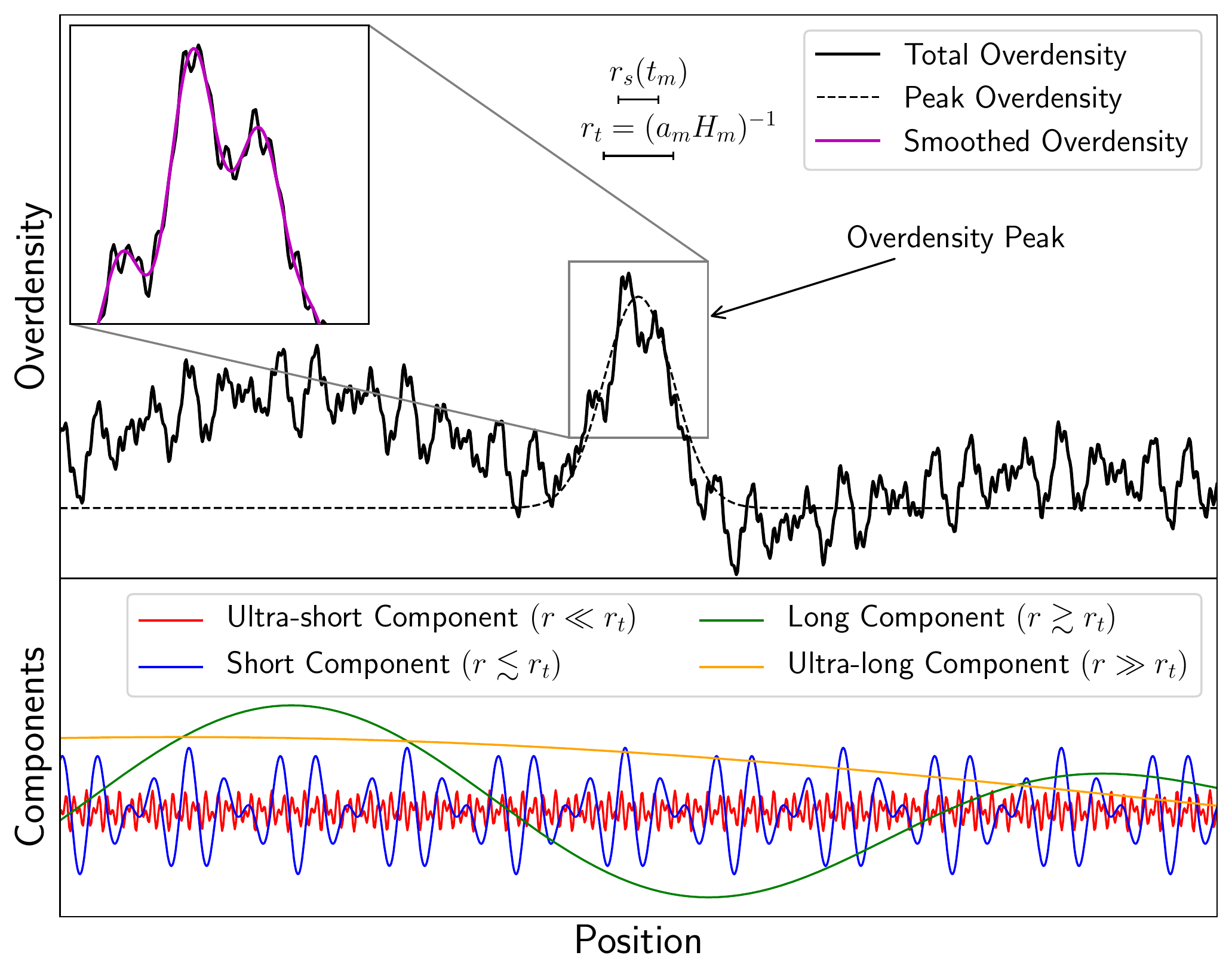}}
\caption{Sketch of the overdensity random field in one spatial dimension. \textit{Upper panel:} total overdensity (\textit{solid line}) given by the sum of a peak overdensity (\textit{dashed line}) with typical scale $r_t$ and the four random fluctuations with different wavelengths illustrated in the lower panel. For completeness we report also the size of the sound horizon $r_s=r_t/\sqrt{3}$. \textit{Zoomed-in panel:} how the peak profile would be before and after smoothing on scale $s$ ($s\ll r_t$, $s\ll r_s$), i.e., removing ultra-short perturbations with frequency $k\gg k_t$. \textit{Lower panel:} the four components of the sketched overdensity with different wavelengths.}
\label{fig:overdensity_field}
\end{figure}

Since the purpose of the filter function in the large-scale structure and CPBHs frameworks is completely different, we cannot apply directly what is typically done in large-scale structure to CPBHs. In particular we have to re-assess the appropriate smoothing scale and how the filtering is performed when the spatial curvature is non-negligible.

To set the smoothing radius, first we have to establish which scales~$r$ (or modes~$k:=r^{-1}$) play a physical role in the collapse. As we have seen in section~\ref{sec:pbh_formation}, we identify the typical scale of a collapsing perturbation in real space with $r_t$, corresponding to a typical mode~$k_t=r_t^{-1}$ in Fourier space\footnote{While in flat space we typically have $k=(2\pi)/r$, in this case, where curvature is non-negligible, the conversion factor between~$k$ and~$r$ will be different from $2\pi$, with a coefficient that depends on the shape. To avoid discussing  scenarios on a case by case basis, we define $k:=1/r$ and we refer the interested reader to Ref.~\cite{yoo:krconversionfactor} for a broader discussion. Notice that the conversion factor is order unity in all the cases of interest, hence this definition does not influence our conclusions.}. To avoid affecting the shape of the peaks, and the properties of the resulting CPBH, the smoothing should be done on scales much smaller than the typical scale of the fluctuation, i.e., for the ``ultra-short'' scales~$r\ll r_t$ or modes~$k\gg k_t$. We show a visual example of an overdensity field and its smoothed version in figure~\ref{fig:overdensity_field}, in particular in the zoomed-in panel. The specific details of how these modes (corresponding to ``ultra-short" scales)  are filtered out should not influence the dynamics of the perturbation on scales~$\mathcal{O}(r_t)$, where the gravitational collapse is the dominant process. Note that this procedure is implicitly implemented every time the spacetime is discretized, as in numerical simulations.

The other relevant physical scale is the size of the sound horizon, where pressure gradients effectively smooth the perturbations. As seen in equation~\eqref{eq:transfer_function}, in linear theory, pressure effects act as an effective filter function that damps perturbations on scales smaller than the sound horizon. At horizon crossing, the size of the sound horizon~$r_s(t_m)=c_s/(a_mH_m)=r_t/\sqrt{3}$ is comparable to the typical scale of the perturbation. Hence, a filter function of smoothing radius~$r_t\equiv 1/(a_mH_m)$ would artificially increase the damping effect produced by pressure. Finally, the height of the peak~$\delta_{\mathrm{peak},0}$, used in equation~\eqref{eq:mass_critical_collapse_II} to determine the mass of the PBH, should not depend on how the filtering procedure is performed (see also appendix~\ref{app:numerical_simulations}).

The condition~$s\ll r_t$ thus~$s\ll r_s(t_m)$, ensures that: \textit{(i)} what determines the true height of the peaks are pressure effects and not the artificial smoothing, \textit{(ii)} does not alter the relevant properties of the field nor the physics involved in CPBH formation and \textit{(iii)} allows us to use all the (cosmological) results about properties of (smoothed) random fields. In other words, the smoothing of the field on small scales should be done on a scale~$r_\mathrm{hor}(t_\mathrm{ini})$, where~$t_\mathrm{ini}$ is some initial time, much smaller than all the typical horizon re-entry time scales~$t_m$.

In the majority of the existing literature the smoothing radius has been typically chosen to be comparable to the typical scale of the perturbation or, equivalently, to the radius of the cosmological comoving horizon when the perturbation starts to collapse, namely~$s\sim\mathcal{O}\left(r_t\equiv 1/a_mH_m\right)$. This operation removes both ultra-short and short perturbation of figure~\ref{fig:overdensity_field}. Some effects of this choice might be seen in Ref.~\cite{ando:pbhsmoothingeffects}, where it was shown that different window functions with this smoothing radius lead to different CPBH abundance constraints. The choice we propose here,~$s\ll r_s(t_m)$, ensures that the details (or even the presence) of the (artificial but unavoidable) smoothing do not affect the dynamic of the collapse and thus the final results including PBH mass and abundance.

It is also important to note that, while the gravitational collapse of a CPBH is a process ongoing at cosmological horizon scales, scales that are well above the horizon at all times during the collapse do not influence sub-horizon dynamics. The collapse is not instantaneous, therefore it is reasonable to expect that long perturbations with modes~$k\lesssim k_t$ might influence the process; on the other hand ultra-long scale perturbations associated to modes~$k\ll k_t$ appear as a constant background during the collapse, hence they should not play any role in determining whether a CPBH forms or not. We refer the reader to figure~\ref{fig:overdensity_field} for a visualization of these two cases. One way to avoid super-horizon effects is to introduce a second (high pass) filter function~$W_{s_2}$ that smooths out scales much larger than the typical scale of the perturbation, i.e., scales~$s_2\gg r_t$ or modes~$k_2\ll k_t$. This is what will be implemented in~\S~\ref{subsec:peak_shape}.

The second difference in the smoothing procedure regards the treatment of non-linearities. So far we have neglected spatial curvature and worked with comoving coordinates. But spatial curvature is not negligible; this introduces some subtleties which ultimately lead to a re-interpretation of the filtering procedure. To account for the curvature, the filtering should be done in physical coordinates~$\mathbf{Y}=ae^{-\zeta(\mathbf{y})}\mathbf{y}$ instead of comoving coordinates~$\mathbf{y}$, as in linear theory. For instance, the filtered overdensity field would be
\begin{equation}
\delta_S(\mathbf{X}) = \int d^3Y W_S(|\mathbf{X}-\mathbf{Y}|)\delta(\mathbf{Y}),
\end{equation}
where $W_S$ is a filter function of smoothing physical scale $S$. Also in this case the filter function has been normalized to unity and it can be written as $W_S=w_S/V^\mathrm{phys.}_w$, where $w_S$ is the unnormalized filter function and $V^\mathrm{phys.}_w=\int d^3Y w_S(|\mathbf{Y}|)$ is the physical volume normalization coefficient. Therefore the overdensity field, filtered on a comoving scale $s$ corresponding to a physical scale $S=ae^{-\zeta(\mathbf{s})}s$, becomes in comoving coordinates
\begin{equation}
\begin{aligned}
\delta_s(\mathbf{x}) &= \delta_{S(s)}\left(\mathbf{X}(\mathbf{x})\right) = \int d^3y e^{-3\zeta(\mathbf{y})} \left[1-\mathbf{y}\cdot\nabla\zeta(\mathbf{y})\right] \frac{a^3}{V^\mathrm{phys.}_w} w_{S(s)}(|\mathbf{X}(\mathbf{x})-\mathbf{Y}(\mathbf{y})|) \delta(\mathbf{y}) \\
&= \int d^3y \frac{a^3}{V^\mathrm{phys.}_w} w_{S(s)}(|\mathbf{X}(\mathbf{x})-\mathbf{Y}(\mathbf{y})|) \delta_\mathrm{new}(\mathbf{y}),
\end{aligned}
\label{eq:correct_smooth overdensity}
\end{equation}
where, using equation~\eqref{eq:overdensity_cosmo_metric}, the ``new'' overdensity field we want to smooth out reads as
\begin{equation}
\delta_\mathrm{new}(\mathbf{y}) = \frac{4}{9}\left(\frac{1}{aH}\right)^2 
e^{-\zeta(\mathbf{y})} \left[1-\mathbf{y}\cdot\nabla\zeta(\mathbf{y})\right] \left[\nabla^2\zeta(\mathbf{y}) - \frac{1}{2}\nabla\zeta(\mathbf{y}) \cdot \nabla\zeta(\mathbf{y}) \right].
\label{eq:overdensity_corrected}
\end{equation}
Notice that in the linear approximation, where curvature and gradients are small ($\zeta\ll 1$, $|\nabla\zeta| \ll 1$), thus $V^\mathrm{phys.}_w\to a^3V^\mathrm{com.}_w$ and $w_S\to w_s$,\footnote{Here we report a practical example of the linear limit. In the case of a Top-Hat filter function in physical coordinates we have $w_S(|\mathbf{Y}|)=\Theta(1-|\mathbf{Y}|/S)$, where $\Theta$ is the Heaviside function. The physical volume reads as $V^\mathrm{phys.}_w = 4\pi S^3/3 = a^3 e^{-3\zeta(\mathbf{s})} 4\pi s^3/3\to a^3V^\mathrm{com.}_w$ when $\zeta\ll 1$. In the same limit the unnormalized filter function reads as $w_S(|\mathbf{Y}|)=\Theta(1-ye^{-\zeta(\mathbf{y})}/se^{-\zeta(\mathbf{s})}) \to \Theta(1-y/s) = w_s(|\mathbf{y}|)$.}, we recover exactly the linear theory definition of filtering of equation~\eqref{eq:linear_filtering_definition}.

However, once the spatial curvature is included in the filtering procedure and the full non-linear relation between overdensity and curvature is used, filtering the overdensity or the curvature fields is not equivalent any more (\textit{cf.} equation~\eqref{eq:smoothed_density_curvature_relation}). Although we do not report the full calculation, it can be easily shown that
\begin{equation}
\begin{aligned}
\delta_s(\mathbf{x}) &= \int d^3y \frac{a^3}{V^\mathrm{phys.}_w} w_{S(s)}(|\mathbf{X}(\mathbf{x})-\mathbf{Y}(\mathbf{y})|) \delta_\mathrm{new}(\mathbf{y}) \\
&\neq \frac{4}{9} \left(\frac{1}{aH}\right)^2 e^{2\zeta_s(\mathbf{x})} \left[\nabla^2\zeta_s(\mathbf{y}) - \frac{1}{2}\nabla\zeta_s(\mathbf{y}) \cdot \nabla\zeta_s(\mathbf{y}) \right],
\end{aligned}
\label{eq:ambiguity_smoothing}
\end{equation}
using the same smoothing function~$w_s$ for both~$\delta$ and~$\zeta$, where the second line of RHS corresponds to equation~\eqref{eq:overdensity_cosmo_metric}. 

Notice also that the curvature enters in the argument of the window function~$a^3W_S$, hence the domain of integration in comoving coordinates will be different from a sphere centred at a given point~$\mathbf{x}$. Large curvature fluctuations will produce large deformation in the domain, however we expect this effect to be negligible when estimating the statistical properties of the whole field since large fluctuations are extremely rare. Moreover, as explained in \S~\ref{subsec:filtering_random_fields}, we are smoothing on scales much smaller than the scale of the peak, therefore we can neglect this curvature dependence when taking correlators of the smoothed overdensity of equation~\eqref{eq:correct_smooth overdensity} and treat the window function as curvature-independent.

According to equation~\eqref{eq:ambiguity_smoothing}, in principle filtering~$\zeta$ is not equivalent to filtering~$\delta$ in the non-linear case. The effect of ignoring this subtlety in the filtering procedures cannot be established a priori, since it depends on the statistical properties of the curvature and density fields. However, given that the abundance of CPBHs depends on those statistical properties, it is important to assess which is the role of the filtering procedure in determining the statistics of the field at initial time (see also \S~\ref{subsec:pbhs_abundance}). Given the general nature of this paper, we leave the quantitative estimation of this effect to a future work. In this paper we will apply the smoothing to the density field.


\subsection{Impact of primordial non-Gaussianities}
\label{subsec:nongaussianity}
Non-Gaussianities can be separated into two categories: primordial non-Gaussianities, generated by some inflationary mechanism and imprinted into the ``matter'' fields at horizon re-entry, and non-Gaussianities generated dynamically, for instance by gravitational evolution. In this section we are interested only in the former. 

The fact that non-linearities are so important in the context of CPBHs already suggests that non-Gaussianities, linked to non-linearities, could be important as well. In fact, primordial non-Gaussianities affect both CPBH formation and abundance and are generally expected to be produced in many of the proposed models for CPBHs generation, see e.g., Refs.~\cite{young:pbhnongaussianity, young:pbhnongaussianityII, young:pbhnongaussianityIII, franciolini:pbhnongaussianity, atal:pbhnongaussianity, passaglia:pbhnongaussianity, kehagias:pbhnongaussianity}. While it has been tested that at cosmological scales initial conditions are very close to be Gaussian~\cite{peiris:wmapng, komatsu:testgaussianity, ade:planckng2013, ade:planckng2015, akrami:planckng2018}, this has not been verified at very small scales, i.e., for the range of scales $k\gtrsim 10^5\ \mathrm{Mpc}^{-1}$ relevant for CPBH formation. 

Here we assess for the first time what is the contribution of primordial non-Gaussianities to the $n$-point function of the smoothed overdensity field including the non-linear effects of equation~\eqref{eq:overdensity_cosmo_metric}. The procedure is exact and can be applied to any $n$-point function, however in this work we concentrate only on the two-point function, i.e., the power spectrum, for reasons that will become clear in section~\ref{sec:peaks_theory}. 

For simplicity, we initially neglect filter and transfer functions and re-introduce them at the end. The full non-linear overdensity field in Fourier is computed from equation~\eqref{eq:overdensity_corrected} yielding:
\begin{equation}
\begin{aligned}
\widehat{\delta}(\mathbf{k}) &= \frac{4}{9}\frac{1}{a^2H^2} \int d^3r \left[1-\mathbf{r}\cdot\nabla\zeta(\mathbf{r})\right] \left[\nabla^2\zeta(\mathbf{r}) - \frac{1}{2}\nabla\zeta(\mathbf{r})\cdot\nabla\zeta(\mathbf{r})\right] e^{-\zeta(\mathbf{r})} e^{-i\mathbf{k}\cdot\mathbf{r}} \\
&= -\frac{4}{9}\frac{1}{a^2H^2} \sum_n \frac{(-1)^n}{n!} \int \frac{d^3q_1}{(2\pi)^3} \frac{d^3q_2}{(2\pi)^3} \frac{d^3q_3}{(2\pi)^3} \left[\widehat{\zeta^0}(\mathbf{q}_1) + 3\widehat{\zeta}(\mathbf{q}_1) - \left.\frac{\partial\widehat{\zeta}(\mathbf{q}_1/\lambda)}{\partial\lambda}\right|_{\lambda=1} \right] \times \\
&\qquad\qquad \times \mathbf{q}_2\cdot\mathbf{q}_3 \left[ \widehat{\zeta}(\mathbf{q}_2) \widehat{\zeta^0}(\mathbf{q}_3-\mathbf{q}_2) \widehat{\zeta^n}(\mathbf{k}-\mathbf{q}_{12}) - \frac{1}{2} \widehat{\zeta}(\mathbf{q}_2)\widehat{\zeta}(\mathbf{q}_3)\widehat{\zeta^n}(\mathbf{k}-\mathbf{q}_{123}) \right],
\end{aligned}
\label{eq:nonlinear_overdensity_fourier_space}
\end{equation}
where we have introduced an auxiliary parameter~$\lambda$,\footnote{Equation~\eqref{eq:nonlinear_overdensity_fourier_space} has been obtained using
\begin{equation*}
\begin{aligned}
\int d^3r \mathbf{r}\cdot\nabla\zeta(\mathbf{r})e^{-i\mathbf{k}\cdot\mathbf{r}} &= \int d^3r \int \frac{d^3q}{(2\pi)^3}i\mathbf{r}\cdot\mathbf{q} \widehat{\zeta}(\mathbf{q}) e^{i\mathbf{q}\cdot\mathbf{r}} e^{-i\mathbf{k}\cdot\mathbf{r}} = \int d^3r \int \frac{d^3q}{(2\pi)^3} \widehat{\zeta}(\mathbf{q}) \left.\frac{\partial e^{i\lambda\mathbf{q}\cdot\mathbf{r}}}{\partial\lambda}\right|_{\lambda=1} e^{-i\mathbf{k}\cdot\mathbf{r}} \\
&= \frac{\partial}{\partial\lambda}\left[ \int d^3r \zeta(\lambda\mathbf{r}) e^{-i\mathbf{k}\cdot\mathbf{r}} \right]_{\lambda=1} = \frac{\partial}{\partial\lambda}\left[\lambda^{-3} \widehat{\zeta}(\mathbf{k}/\lambda) \right]_{\lambda=1} = -3\widehat{\zeta}(\mathbf{k}) + \left.\frac{\partial\widehat{\zeta}(\mathbf{k}/\lambda)}{\partial\lambda}\right|_{\lambda=1}.
\end{aligned}
\end{equation*}} and we have expanded the exponential~$e^{-\zeta}$ in series, introducing the function~$\widehat{\zeta^n}$, which is the Fourier transform of~$\zeta^n$. The function~$\widehat{\zeta^n}$ can be computed at every order in $n$ and it is given by
\begin{equation}
\begin{aligned}
\widehat{\zeta^0}(\mathbf{k}^\star) &= (2\pi)^3\delta^{D}(\mathbf{k}^\star)  &n=0,\\
\widehat{\zeta^1}(\mathbf{k}^\star) &= \widehat{\zeta}(\mathbf{k}^\star) &n=1,\\
\widehat{\zeta^n}(\mathbf{k}^\star) &= \int\prod_{j=1}^{n-1}\left[\frac{\mathrm{d^3}p_j}{(2\pi)^3}\widehat{\zeta}(\mathbf{p}_j)\right] \widehat{\zeta}\left(\mathbf{k}^\star - \sum_{j=1}^{n-1}\mathbf{p}_j \right)  &n\geq 2.
\end{aligned}
\label{eq:zeta_n_fourier}
\end{equation}
Because of non-linearities, even when the $\zeta$-curvature field is Gaussian and has zero one-point correlator $\left(\langle\widehat{\zeta}(\mathbf{k}) \rangle=0\right)$, the one-point correlator of the overdensity field is non-zero. Its exact value can be computed from equation~\eqref{eq:nonlinear_overdensity_fourier_space} and it can be checked that it is non-zero only for the ultra-long mode $\mathbf{k}=0$ and it depends only on the statistical properties of the field:
\begin{equation}
\left\langle\widehat{\delta}(\mathbf{k})\right\rangle = (2\pi)^3\delta^D(\mathbf{k}) \mathcal{G}(\sigma^2_j),
\end{equation}
where the $\mathcal{G}(\sigma^2_j)$ is a function of the spectral moments of the $\zeta$-curvature field (see section~\ref{sec:peaks_theory} for the definition of spectral moments).

Notice that in computing the two-point function $\left\langle\widehat{\delta}(\mathbf{k}_1)\widehat{\delta}(\mathbf{k}_2)\right\rangle$ of the overdensity field we should include also the disconnected component $\left\langle\widehat{\delta}(\mathbf{k}_1)\right\rangle\left\langle\widehat{\delta}(\mathbf{k}_2)\right\rangle$, however this contribution is zero for our range of scales of interest, i.e., when $\mathbf{k}\neq 0$. On the other hand, the connected component of the overdensity field power spectrum is given by
\begin{equation}
\begin{aligned}
& \left\langle\widehat{\delta}(\mathbf{k}_1)\widehat{\delta}(\mathbf{k}_2)\right\rangle_\mathrm{c} = \frac{16}{81} \left(\frac{1}{aH}\right)^4 \sum_{n,m} \frac{(-1)^{n+m}}{n!m!} \int \frac{d^3q_1}{(2\pi)^3} \frac{d^3q_2}{(2\pi)^3} \frac{d^3q_3}{(2\pi)^3} \frac{d^3q_4}{(2\pi)^3} \frac{d^3q_5}{(2\pi)^3} \frac{d^3q_6}{(2\pi)^3} (\mathbf{q}_2\cdot\mathbf{q}_3) (\mathbf{q}_5\cdot\mathbf{q}_6) \times  \\
& \times \left\langle \left[\widehat{\zeta^0}(\mathbf{q}_1) + 3\widehat{\zeta}(\mathbf{q}_1) - \left.\frac{\partial\widehat{\zeta}(\mathbf{q}_1/\lambda)}{\partial\lambda}\right|_{\lambda=1} \right] \left[ \widehat{\zeta}(\mathbf{q}_2) \widehat{\zeta^0}(\mathbf{q}_3-\mathbf{q}_2) \widehat{\zeta^n}(\mathbf{k}_1-\mathbf{q}_{12}) - \frac{1}{2} \widehat{\zeta}(\mathbf{q}_2)\widehat{\zeta}(\mathbf{q}_3)\widehat{\zeta^n}(\mathbf{k}_1-\mathbf{q}_{123}) \right] \right. \\
& \left. \left[\widehat{\zeta^0}(\mathbf{q}_4) + 3\widehat{\zeta}(\mathbf{q}_4) - \left.\frac{\partial\widehat{\zeta}(\mathbf{q}_4/\lambda)}{\partial\lambda}\right|_{\lambda=1} \right] \left[ \widehat{\zeta}(\mathbf{q}_5) \widehat{\zeta^0}(\mathbf{q}_6-\mathbf{q}_5) \widehat{\zeta^m}(\mathbf{k}_2-\mathbf{q}_{45}) - \frac{1}{2} \widehat{\zeta}(\mathbf{q}_5)\widehat{\zeta}(\mathbf{q}_6)\widehat{\zeta^m}(\mathbf{k}_2-\mathbf{q}_{456}) \right] \right\rangle
\end{aligned}
\label{eq:two_point_function_fourier}
\end{equation}
First of all we stress that this result is exact, no approximation has been taken so far. Second, we see that the entire family of $n$-point function of the $\zeta$-curvature contributes to the two-point function of the overdensity due to the $e^{-\zeta}$ factor. In this sense, non-linearities and primordial non-Gaussianities are very important in determining the full power-spectrum of the overdensity field. From the inflationary model-building side, it is therefore important not only to compute the primordial curvature power spectrum, but also the primordial bispectrum and  higher-order correlations, and to assess the magnitude of their contribution, given by equation~\eqref{eq:two_point_function_fourier}, to the overdensity two-point function. 

As we will explain in section~\ref{sec:peaks_theory}, we are interested in filtering only the overdensity field, hence filter functions can be introduced multiplying equation~\eqref{eq:two_point_function_fourier} by $\widehat{W'}_s(\mathbf{k}_1)\widehat{W'}_s(\mathbf{k}_1)$, where $\widehat{W'}_s$ is the Fourier transform of the filter function $a^3W_S$. As explained after equation~\eqref{eq:ambiguity_smoothing}, this procedure is not strictly correct, since the curvature perturbation appears also in the argument of the filter function, deforming the volume of the smoothing region. However, since we are performing the smoothing on scales much smaller than the scale of the perturbations, this effect is expected to be sub-leading. The other (physical) smoothing is the one introduced by the sound horizon and described by the transfer function. The transfer function, by definition describes the evolution of the curvature perturbation on sub-horizon modes, hence in equations~\eqref{eq:zeta_n_fourier} and~\eqref{eq:two_point_function_fourier}, every curvature perturbation~$\widehat{\zeta}$ should be substituted with $\widehat{\mathcal{T}}_\mathrm{NL}\widehat{\zeta}$, where $\widehat{\mathcal{T}}_\mathrm{NL}$ is the fully non-linear transfer function.

Finally, we provide a practical example of how equation~\eqref{eq:two_point_function_fourier} can be used to compute the leading bispectrum~$B_\zeta$ contribution to the two-point function of the smoothed overdensity field:
\begin{equation}
\begin{aligned}
& P_s (\tau, k) = \frac{16}{81} \frac{k^4}{a^4H^4} \widehat{W'}^2_s(k) \widehat{\mathcal{T}}^2_\mathrm{NL}(\tau,k) \Bigg[ P_\zeta(k) \ +  \frac{1}{k^2\widehat{\mathcal{T}}_\mathrm{NL}(\tau,k)}\times \\
& \left.  \int \frac{d^3q}{(2\pi)^3} \widehat{\mathcal{T}}_\mathrm{NL}(\tau,q) \widehat{\mathcal{T}}_\mathrm{NL}(\tau,|\mathbf{k}+\mathbf{q}|) \left[6|\mathbf{k}+\mathbf{q}|^2-2q^2 + \mathbf{q}\cdot(\mathbf{k}+\mathbf{q})\right]B_\zeta(k,q,|\mathbf{k}+\mathbf{q}|) + \cdots \right].
\end{aligned}
\label{eq:nongaussian_smoothed_power_spectrum}
\end{equation}
The dots in the second line represent sub-leading Gaussian and mixed Gaussian/non-Gaussian corrections, generated for instance by terms proportionals to $\left\langle\zeta\zeta\zeta\zeta\right\rangle\propto P_\zeta P_\zeta$ and $\left\langle\zeta\zeta\zeta\zeta\zeta\right\rangle\propto P_\zeta B_\zeta$, respectively\footnote{The fact that these extra Gaussian terms are subleading can be easily seen by noticing that they contain four transfer functions instead of just two, as in the leading term, hence they will be highly suppressed compared to the first line of equation~\eqref{eq:nongaussian_smoothed_power_spectrum}. The same reasoning applies also to mixed terms, where there will be even more transfer functions.}. Similar terms appear also in the analysis of clustering properties of halos, see e.g., Ref.~\cite{bellomo:gravitonexchange}, and they appear to be a general prediction of non-linear overdensity fields.

Equation~\eqref{eq:nongaussian_smoothed_power_spectrum} reduces to the well-known form of equation~\eqref{eq:smoothed_overdensity_Pk} (but with a different window function) in case of Gaussian initial conditions, i.e., $B_\zeta\equiv 0$, and neglecting sub-leading contributions. This equation also shows that the relation between the power spectrum of $\delta$ and that of $\zeta$ is not univocal: in this example, the same smoothed overdensity power spectrum can be generated by a non-Gaussian $\zeta$ with power spectrum $P_\zeta$ and bispectrum $B_\zeta$ or by a Gaussian $\zeta'$ with power spectrum $P_{\zeta'}$ equal to the argument of the square bracket in the RHS of equation~\eqref{eq:nongaussian_smoothed_power_spectrum}. We conclude that particular care is needed when trying to reconstruct the primordial curvature power spectrum from the overdensity two-point function (see section~\ref{sec:primordial_power_spectrum_reconstruction}).


\section{Peak theory applied to Primordial Black Holes}
\label{sec:peaks_theory}
Since CPBHs form from peaks in the overdensity field, we resort to peak theory~\cite{bardeen:peakstheory} to calculate PBH abundance at formation time, starting from the statistical distribution of the local maxima. In order to define a maxima, we need a smooth and differentiable random field, obtained by using the prescription given in \S~\ref{subsec:filtering_random_fields}. In principle the field can be either Gaussian or non-Gaussian, however we work with a smoothed Gaussian random overdensity field, whose statistical properties are completely specified by its power spectrum~$P_s$ or, alternatively, by its Fourier transform, the correlation function~$\xi_s(|\mathbf{x}_1-\mathbf{x}_2|)=\left\langle\delta_s(\mathbf{x}_1)\delta_s(\mathbf{x}_2)\right\rangle$. To be accurate, the Gaussian assumption might be too strict for the PBHs scenario, in fact we already saw in \S~\ref{subsec:nongaussianity} that non-linearities induce departures from the Gaussian statistics in the overdensity field. However peak theory has not been fully developed for non-Gaussian fields, hence we will consider only the Gaussian case.

The original framework considers the statistics of a smoothed random field deep in the matter-dominated era and it connects the initial statistics to the formation of large-scale structure. The  original framework was designed for matter-domination, where pressure effects are negligible, hence we need to slightly upgrade it to employ it also during radiation-domination. In fact, pressure effects on sub-horizon scales naturally wash out perturbations, changing the statistical properties of the random field. 


\subsection{Primordial Black Holes abundance}
\label{subsec:pbhs_abundance}
The statistical properties of the field are encoded in the spectral moments
\begin{equation}
\sigma^2_j(\tau) = \int \frac{d^3k}{(2\pi)^3} k^{2j} P_s(k,\tau),
\label{eq:spectral_moments}
\end{equation}
where the smoothed power spectrum is given in equation~\eqref{eq:nongaussian_smoothed_power_spectrum} after imposing Gaussian initial conditions ($B_\zeta\equiv 0$). 

The statistical properties must be evaluated for the entire field at once, therefore in this case we are not filtering out ultra-long scales (\textit{cf.} \S~\ref{subsec:filtering_random_fields}). In section~\ref{sec:pbh_formation} we considered each perturbation separately; in reality the Universe is filled by a superposition of perturbations, with a distribution of shapes (or, equivalently,  of shape parameter~$\alpha$) and typical scales. If PBHs form from rare events drawn from the tail of the probability distribution of peaks, at any given time the distribution of~$\alpha$ and typical scales for PBH ``seeds'' is likely to be fairly narrow. In particular, the typical scale is of the order of the horizon and the higher the peak, the narrower the distribution of~$\alpha$ (see also \S~\ref{subsec:peak_shape}).

As anticipated in \S~\ref{subsec:filtering_random_fields}, the smoothing of the field on small scales should be done on a scale~$r_\mathrm{hor}(\tau_\mathrm{ini})$, where~$\tau_\mathrm{ini}$ is some initial time, much smaller than all the typical scales of the entire set of perturbations that will be relevant for PBH formation. The exact value of~$\tau_\mathrm{ini}$ is not relevant, as long as the condition is satisfied. The smoothing procedure should also guarantee that the three spectral moments~$\sigma_0$, $\sigma_1$ and~$\sigma_2$ are finite at initial time. This extra requirement ensures that spectral moments remain finite at every time and that statistical properties are well-defined. Any result of peak theory cannot be used unless this preliminary requisite is satisfied. Note that if statistical properties are evaluated incorrectly because of a wrong filtering procedure, one could reach incorrect conclusions  and thus incorrect inference of  cosmology.

Once spectral moments are well defined, we can construct the spectral parameters
\begin{equation}
\gamma(\tau) = \frac{\sigma_1^2(\tau)}{\sigma_2(\tau)\sigma_0(\tau)}, \qquad R_\star(\tau) = \sqrt{3}\frac{\sigma_1(\tau)}{\sigma_2(\tau)},
\label{eq:spectral_parameters}
\end{equation}
which completely determine the comoving density of peaks. In fact, by defining the relative peak height as~$\nu=\delta_\mathrm{peak,0}/\sigma_0$, the differential comoving peak density reads~\cite{bardeen:peakstheory}
\begin{equation}
\frac{dn^\mathrm{com.}_\mathrm{peak}(\tau,\nu)}{d\nu} = \frac{1}{(2\pi)^2R^3_\star(\tau)}e^{-\nu^2/2} G\left(\gamma(\tau),\gamma(\tau)\nu\right),
\label{eq:comoving_peak_number_density}
\end{equation}
where the function~$G(\gamma,\gamma\nu)$ can be approximated by~\cite{bardeen:peakstheory}
\begin{equation}
G(\gamma,\omega) \approx \frac{\omega^3-3\gamma^2\omega+[B(\gamma)\omega^2+C_1(\gamma)]e^{-A(\gamma)\omega^2}}{1+C_2(\gamma)e^{-C_3(\gamma)\omega}}
\label{eq:G_definition}
\end{equation}
for~$0.3<\gamma<0.7$ and for~$-1<\omega<+\infty$, keeping the difference between the numerical result and the analytic estimation below~$1\%$\footnote{The coefficients of equation~\eqref{eq:G_definition} are given by~\cite{bardeen:peakstheory}
\begin{equation*}
\begin{aligned}
A(\gamma) = \frac{5/2}{9-5\gamma^2},\qquad & B(\gamma) = \frac{432}{\left(10\pi\right)^{1/2}\left(9-5\gamma^2\right)^{5/2}},	\\
C_1(\gamma) = 1.84 + 1.13\left(1-\gamma^2\right)^{5.72}, \quad &C_2(\gamma) = 8.91 + 1.27e^{6.51\gamma^2}, \quad C_3(\gamma) = 2.58e^{1.05\gamma^2}.
\end{aligned}
\end{equation*}}. 

During matter-dominated era, at linear order, the overdensity field grows self-similarly on every scale, hence all the spectral moments share the same time dependencies and the spectral parameters of equation~\eqref{eq:spectral_parameters} are time-independent. Therefore, in the large-scale structure framework, the comoving density of peaks does not depend on the time at which it is computed.  This is not the case during radiation-domination where the magnitude of the spectral moments diminishes with time because of the suppression of perturbations with high wave modes~$k$ due to pressure effects (see equation~\eqref{eq:transfer_function}). Therefore, evaluating equation~\eqref{eq:comoving_peak_number_density} at different times yield different number densities because the random field itself is different.

The evolution of the field is uniquely determined by the cosmic expansion and pressure effects, therefore it is possible to compute consistently the comoving peak density at any time by accounting for these two  effects. For convenience we choose to evaluate the differential physical number density of peaks at horizon crossing (conformal) time (see Sec \ref{subsec:formation_criterion}), $\tau_m$:
\begin{equation}
\frac{dn^\mathrm{phys.}_\mathrm{peak}(\nu,\tau_m)}{d\nu} = a^{-3}_m\frac{dn^\mathrm{com.}_\mathrm{peak}(\nu,\tau_m)}{d\nu}.
\end{equation}
This is the number density of regions that will collapse and form a PBH at formation time~$\tau_f\gtrsim\tau_m$. Notice that, since CPBH formation is not instantaneous, the differential physical number density of CPBHs has to be rescaled by a factor~$a^{-3}_f/a^{-3}_m$, where~$a_f=a(\tau_f)$. Numerical simulations show that the the cosmic time of formation~$t_f / t_m \simeq 10$ \cite{musco:pbhthreshold}, where~$t_m$ is the horizon crossing time defined in \S~\ref{subsec:formation_criterion}, and therefore~$\tau_f/\tau_m = a_f/a_m \simeq 3$. The exact relation between horizon crossing and formation time is not important, since this factor cancels out when computing the PBHs abundance today (see section~\ref{sec:primordial_power_spectrum_reconstruction}, equation~\eqref{eq:sigma0_equation}).

We have already discussed in section~\ref{sec:pbh_formation} that not all the peaks correspond to sites where a CPBH forms and that we need a threshold criterion to assess which peaks collapse and which do not. As seen in equations~\eqref{eq:mass_critical_collapse} and~\eqref{eq:mass_critical_collapse_II}, in this case the criterion to start gravitational collapse reads as $\delta_{\mathrm{peak},0}(\tau_m)>\delta_{\mathrm{peak},0,c}(\alpha)$, where $\delta_{\mathrm{peak},0,c}(\alpha)$ depends on the shape of the perturbation. Since CPBHs are non-relativistic compact objects, their differential energy density is expected to be written as $d\rho_\mathrm{PBH}/d\nu \propto M_\mathrm{PBH}(\nu) dn_\mathrm{peak}/d\nu$. Here there are two subtleties that enter in $M_\mathrm{PBH}(\nu)$: the time evolution of $\delta_{\mathrm{peak},0}(\tau_m)$ and the shape dependence of the threshold. 

Two considerations are in order. First, the amplitude of the peak,~$\delta_{\mathrm{peak},0}$, has been computed considering only cosmic expansion and not pressure effects. Thus the argument of~$M_\mathrm{PBH}$ should be~$\nu'$ instead of~$\nu$, because it refers to the relative amplitude of a field that has been evolved in time in a different way. On the other hand, if we evaluate the effects of pressure, i.e., the contribution due the linear transfer function, on scales close to the typical scale of the perturbation, i.e., for~$k\tau\simeq 1$, we find that~$\widehat{\mathcal{T}}_\mathrm{LIN}\simeq 0.9$. However the required transfer function in our case is the non-linear one~$\widehat{\mathcal{T}}_\mathrm{NL}$. It is not possible to estimate the non-linear transfer function from the linear one reported in equation~\eqref{eq:transfer_function}. It has been shown for instance in Refs.~\cite{bartolo:secondordertransfer, bartolo:secondordertransferII, bartolo:secondordertransferIII}, that the transfer function at second order in perturbation theory is not simply the square of the linear transfer function, as one would have na\"ively expected. We leave the derivation of the non-linear transfer function for future work~\cite{byrnes:nonlineartransfer}, however here we give a first estimate of these effects. By comparing the profiles obtained assuming linear theory, i.e., using the linear transfer function, and profiles obtained from our numerical simulations, we estimate the relative difference between the non-linear and linear transfer function as the average difference between the density profiles for scales smaller than the typical scale of the perturbation:
\begin{equation}
\frac{\left[\left(\widehat{\mathcal{T}}_\mathrm{NL}-\widehat{\mathcal{T}}_\mathrm{LIN}\right)^2\right]^{1/2}}{\widehat{\mathcal{T}}_\mathrm{LIN}} \sim \frac{\left[\left(\bar{\delta}^\mathrm{NL}_\mathrm{peak}-\bar{\delta}^\mathrm{LIN}_\mathrm{peak}\right)^2\right]^{1/2}}{\bar{\delta}^\mathrm{LIN}_\mathrm{peak}},
\label{eq:correction_transfer_function}
\end{equation}
where the profiles $\bar{\delta}^\mathrm{NL}_\mathrm{peak}$ have been obtained using numerical simulations. Even if the estimation is crude, we find that $\widehat{\mathcal{T}}_\mathrm{NL}/\widehat{\mathcal{T}}_\mathrm{LIN} \simeq 1.8,\ 1.5,\ 1.2$ for profiles characterised by~$\alpha=0.15,\ 1.0,\ 30.0$, respectively. In conclusion we find that at horizon crossing, pressure effects are inefficient at smoothing inhomogeneities on horizon scales at horizon re-entry time simply because such region was not in causal contact before and effects due to the non-linear growth of the perturbation have not produce significant deviations from the linear theory. Therefore, in what follows, we will assume~$\nu'\simeq \nu$ for simplicity.

We also find that the position of the maximum of the compaction function changes by a factor~$10-15\%$ towards larger values with respect to the position estimated using initial conditions (\textit{cf.} section~\ref{sec:pbh_formation}). Therefore the true horizon crossing happens later than what predicted by initial conditions criteria, even when considering non-linearities (see the updated version of Ref.~\cite{musco:pbhthreshold} for more details). This difference in horizon crossing scales generates a~$20\%$ difference in the horizon crossing time and in the mass contained inside the horizon, however these differences will not affect significantly our conclusions because what really determines the CPBH mass is how much above the critical threshold the perturbation is.

Second, the threshold is shape-dependent, therefore the approach to get the correct energy density would be to integrate 
\begin{equation}
\frac{d^{10}\rho_\mathrm{PBH}}{d\nu d^3\mathcal{J} d^6\mathcal{H}} \propto M_\mathrm{PBH}(\nu, \bm{\mathcal{J}}, \bm{\mathcal{H}}) \frac{d^{10}n_\mathrm{peak}}{d\nu d^3\mathcal{J} d^6\mathcal{H}},
\end{equation}
where $\bm{\mathcal{J}}$ is a three-dimensional vector containing information on first derivatives ($\eta_j$, with $j\in\{1,2,3\}$ in the notation of Ref.~\cite{bardeen:peakstheory}) and $\bm{\mathcal{H}}$ is a six-dimensional vector containing information on second derivatives ($\zeta_{ij}$, with $ij\in\{11,22,33,12,13,23\}$ in Ref.~\cite{bardeen:peakstheory}). Since the estimation of this integral for a population of different perturbations with different shapes goes beyond the purpose of this paper, in the following we assume that all the peaks share the same shape, i.e.,~$M_\mathrm{PBH}(\nu, \bm{\mathcal{J}}, \bm{\mathcal{H}})=M_\mathrm{PBH}(\nu)$, thus the same (time-dependent) critical threshold~$\nu^{(\alpha)}_{c,m} = \delta_{\mathrm{peak},0,c}(\alpha)/\sigma_0(\tau_m)$.

Therefore, under these approximation and following Ref.~\cite{germani:pbh_abundance}, we define the relative energy density of CPBHs with respect to the energy density of radiation at formation time as
\begin{equation}
\begin{aligned}
\beta(\tau_f) &= \frac{\rho_\mathrm{PBH}(\tau_f)}{\rho_\mathrm{rad}(\tau_f)} = \frac{1}{\rho_\mathrm{rad}(\tau_f)}\int_{\nu_{c,m}^{(\alpha)}}^{\infty} d\nu \frac{d\rho^\mathrm{phys.}_\mathrm{PBH}(\nu,\tau_f)}{d\nu} \\
&= \frac{1}{\rho_\mathrm{rad}(\tau_f)}\int_{\nu_{c,m}^{(\alpha)}}^{\infty} d\nu \left(\frac{a_f}{a_m}\right)^{-3} M_\mathrm{PBH}(\nu,\tau_m) \frac{dn^\mathrm{phys.}_\mathrm{peak}(\nu,\tau_m)}{d\nu}.
\end{aligned}
\label{eq:pbh_abundance}
\end{equation}
Given that CPBHs effectively behave as dark matter, we can connect the abundance at initial time to the constrained abundance~$f_\mathrm{PBH}=\rho_\mathrm{PBH}/\rho_\mathrm{dm}$, i.e., the fraction of dark matter in PBHs (assuming that they all form at the time~$\tau_f$), as~\cite{nakama:gstar, nakama:gstar2}
\begin{equation}
\beta(\tau_f) = \frac{\rho_\mathrm{PBH}(\tau_f)}{\rho_\mathrm{rad}(\tau_f)} = \frac{g_{\star, \rho}(\tau_0)}{g_{\star, \rho}(\tau_f)} \left(\frac{g_{\star, s}(\tau_f)}{g_{\star, s}(\tau_0)}\right)^{4/3} \frac{\Omega_{\mathrm{dm},0}}{\Omega_{\mathrm{rad},0}} f_\mathrm{PBH} a_f,
\label{eq:pbh_abundance_II}
\end{equation}
where~$\tau_0$ is the conformal time today, $\Omega_{\mathrm{dm},0}$ and $\Omega_{\mathrm{rad},0}$ are the present dark matter and radiation densities with respect to the critical density today~$\rho_{0c}$, while~$g_{\star, \rho}$ and~$g_{\star, s}$ are the total number of effective degrees of freedom for the energy density and the entropy density~\cite{husdal:degreesoffreedom}. Significantly different approximations of equation~\eqref{eq:pbh_abundance_II} have been used in the literature; we report here the correct result and we refer the reader to appendix~\ref{app:degrees_of_freedom} for its derivation, our choice of  values for the effective degrees of freedom  and further comments, especially on the role of neutrinos. Typical values of initial abundance are~$\beta(\tau_f)\simeq 10^{-17}$ ($\beta(\tau_f)\simeq 10^{-6}$) for~$M_\mathrm{hor}\simeq 10^{-18}\ M_\odot$ ($M_\mathrm{hor}\simeq 10^{4}\ M_\odot$) and~$f_\mathrm{PBH}=1$, explicitly showing that regions where overdensity perturbations are large enough to collapse are very rare (hence justifying also  our treatment of windows functions in \S~\ref{subsec:filtering_random_fields} and~\S~\ref{subsec:nongaussianity}).

The definition in equation~\eqref{eq:pbh_abundance} is accurate only when all the CPBHs form at a given time (or, equivalently, at a given scale). However, in a realistic scenario, CPBHs form over some time interval, therefore equations~\eqref{eq:pbh_abundance} and~\eqref{eq:pbh_abundance_II} should be interpreted as $d\beta/d\tau_f$ and the total abundance of CPBHs would become
\begin{equation}
\beta_f^\mathrm{tot} = \int_{\tau_\mathrm{min}}^{\tau_\mathrm{max}} d\tau_f \frac{d\beta}{d\tau_f},
\end{equation}
with the condition~$\tau_m=\tau_f/3$, $\tau_\mathrm{min}\gg \tau_\mathrm{ini}$ to avoid being biased by the filtering procedure, and~$\tau_\mathrm{max}\lesssim \tau_\mathrm{eq}$ since we are interested only in PBHs forming during the radiation-dominated era.

In principle peak theory would suffer from the cloud-in-cloud problem or, in this case, the black-hole-in-black-hole problem. If CPBHs are generated by a localised peak in the primordial power spectrum, this problem is unimportant. For a  very broad peak or a plateau, it may in principle be an issue but since CPBHs are exceedingly rare the black-hole-in-black-hole problem should be much reduced compared to the standard  cloud-in-cloud one for dark matter haloes (see also e.g., Ref.~\cite{moradinezhad:pbhclustering}). In the context of large-scale structure this problem has already been cured in Ref.~\cite{paranjape:excursionpeaktheory}, introducing the so called Excursion Set Peaks formalism. We leave its implementation to future work.


\subsection{The shape of the overdensity peak}
\label{subsec:peak_shape}
Peak theory enables us to connect the average shape of the peaks to the statistical properties of the random field. Assuming that at~$\mathbf{x}=0$ there is a peak of height~$\nu$ and averaging over all possible curvatures and orientations\footnote{In principle is not mandatory to average over curvatures and orientations, however deviations from spherical symmetry are suppressed by a factor~$1/\nu$~\cite{bardeen:peakstheory}, hence they are suppressed for high peaks, as in the cases of interest. Moreover the numerical simulations we use assume spherical symmetry, therefore our choice is natural.}, the mean value of the overdensity at distance~$r$ from the peak, i.e., the average shape of the peak, can be written as~\cite{bardeen:peakstheory} 
\begin{equation}
\frac{\bar{\delta}_\mathrm{peak}(\tau,r)}{\sigma_{0}(\tau)} = \nu\psi(\tau,r) - \frac{\theta\left(\gamma(\tau),\gamma(\tau)\nu\right)}{\gamma(\tau)\left(1-\gamma^2(\tau)\right)}\left[ \gamma^2(\tau)\psi(\tau,r) + \frac{\nabla^2\psi(\tau,r)}{3} \right],
\label{eq:peak_average_shape}
\end{equation}
where $\psi(\tau, r) = \xi_s(\tau, r)/\sigma_0^2(\tau)$ and the function $\theta(\gamma,\omega)$ is given by~\cite{bardeen:peakstheory}
\begin{equation}
\theta(\gamma,\omega) = \frac{3(1-\gamma^2) + (1.216-0.9\gamma^4)e^{-2\gamma/\omega^2}}{\left[3(1-\gamma^2) + 0.45 + \frac{\omega^2}{4} \right]^{1/2} + \frac{\omega}{2}}
\end{equation}
and it is accurate for~$\gamma\in[0.4,0.7]$ and~$\omega\in[1,3]$. This result is very similar to the average density profile around a point with the same height~$\nu$ as the peak but which is not a peak, in fact in the latter case we have~\cite{rice:correlationprofiles, dekel:correlationprofiles}
\begin{equation}
\frac{\bar{\delta}_\mathrm{no-peak}(\tau,r)}{\sigma_{0}(\tau)} = \nu\psi(\tau,r),
\label{eq:nopeak_average_shape}
\end{equation}
where equations~\eqref{eq:peak_average_shape} and~\eqref{eq:nopeak_average_shape} coincide in the limit of high $\nu$, since for high thresholds virtually all regions are peaks.

Not all peaks share the same shape, therefore we associate a variance of shapes~$\sigma^2_\mathrm{peak}(\tau, r)$ and~$\sigma^2_\mathrm{no-peak}(\tau, r)$ to the mean profiles of equations~\eqref{eq:peak_average_shape} and~\eqref{eq:nopeak_average_shape} (see Ref.~\cite{bardeen:peakstheory} for the explicit form of the variance). For high peaks the variance is small, however far from the peak the variance grows and it becomes as large as the amplitude of the overdensity itself. Following Ref.~\cite{dekel:correlationprofiles}, we call this distance the decoherence distance~$r_\mathrm{dec}$, because at this point we cannot distinguish any more if we are ``in a peak'' or not. In the limit of zero shape variance ($\sigma^2_\mathrm{peak},\sigma^2_\mathrm{no-peak}\equiv 0$) the decoherence distance corresponds to the zero-crossing distance~$r_0$ defined in section~\ref{sec:pbh_formation} for the family of profiles under consideration. A more conservative choice of decoherence distance is given by~$r_m$, however the difference between~$r_0$ and~$r_m$ is of~$\mathcal{O}(\mathrm{1})$ in our cases, therefore choosing one or the other does not significantly affect our results\footnote{In general the ratio~$r_0/r_m$ varies from~$1$ to~$\infty$, however it has been shown that shapes with a similar behaviour in the region~$r\lesssim r_m$, but a different one in the outward region, have almost the same threshold. The variation is at most few percent even when~$r_0/r_m$ changing significantly~\cite{musco:pbhthreshold}.}.

Since we consider high peaks (or alternatively, rare events), we can neglect the $\theta (\gamma,\gamma\nu)$ correction in equation~\eqref{eq:peak_average_shape} and use
\begin{equation}
\xi_s(\tau,r) = \sigma^2_{0}(\tau) \frac{\bar{\delta}_\mathrm{peak}(\tau,r)}{\bar{\delta}_\mathrm{peak}(\tau,0)},
\label{eq:smoothed_correlation_function_delta_peaks}
\end{equation}
which is valid for scales smaller than the decoherence radius. For distances greater than the decoherence length, the density fluctuations become ``uncorrelated''. In this regime, an estimate of the two-point correlation function~$\Xi_s(\tau, r)=\xi_s(\tau, r>r_\mathrm{dec})$ can be obtained, for instance, by studying the effects of primordial clustering of PBHs~\cite{chisholm:pbhclusteringI, chisholm:pbhclusteringII, alihaimoud:pbhclustering, desjacques:pbhclustering, ballesteros:pbhclustering, moradinezhad:pbhclustering, garriga:pbhclustering}, however this goes beyond the scope of this article, hence we leave it for future work.

Equivalently, one can also work with the Fourier transform of equation~\eqref{eq:smoothed_correlation_function_delta_peaks}, i.e., the smoothed power spectrum, which reads as
\begin{equation}
\begin{aligned}
P_s(\tau, k) &=  \int d^3r \xi_s(\tau, r) e^{-i\mathbf{k}\cdot\mathbf{r}} \\
&= 4\pi \left[\sigma^2_{0}(\tau) \int_{0}^{r_\mathrm{dec}} dr\ r^2 \frac{\sin(kr)}{kr}\frac{\bar{\delta}_\mathrm{peak}(\tau, r)}{\bar{\delta}_\mathrm{peak}(\tau, 0)} + \int_{r_\mathrm{dec}}^\infty dr\ r^2 \frac{\sin(kr)}{kr}\Xi_s(\tau, r)\right].
\end{aligned}
\label{eq:reconstructed_power_spectrum}
\end{equation}
Here we can neglect the second integral in the second line of equation~\eqref{eq:reconstructed_power_spectrum} because of the~$\sin(x)/x$ suppression factor, as long as we consider modes~$k\gtrsim k_\mathrm{dec} = r^{-1}_\mathrm{dec}$, i.e., modes that play a role in the gravitational collapse. Moreover, all those scales are super-horizon at horizon crossing time, hence they should be filtered out using the second window function defined in \S~\ref{subsec:filtering_random_fields}.  Equation~\ref{eq:reconstructed_power_spectrum} makes  evident how an incorrect estimate of the profile of the peak yields a mis-estimation of the statistical properties of the field. For this reason, in this work we have used a family of profiles which covers multiple possibilities. In section~\ref{sec:primordial_power_spectrum_reconstruction} we report the constraints on the power spectrum obtained from the entire family.


\section{The reconstruction of primordial power spectrum amplitude and shape}
\label{sec:primordial_power_spectrum_reconstruction}
We now combine the results from  the previous sections, our three pillars, to reconstruct both the amplitude and the shape of the primordial curvature power spectrum, assuming Gaussian initial conditions.

By combining equation~\eqref{eq:nongaussian_smoothed_power_spectrum} (in the Gaussian limit,~$B_\zeta\equiv 0$) and equation~\eqref{eq:reconstructed_power_spectrum}, and evaluating both at horizon crossing, we obtain
\begin{equation}
\begin{aligned}
\widehat{W'}^2_s(k) \widehat{\mathcal{T}}_\mathrm{NL}^2(\tau_m,k) P_\zeta(k) = \frac{81}{16} \left(\frac{a_m H_m}{k}\right)^4 \times 4\pi\sigma_0^2(\tau_m) \int_{0}^{r_t} dr\ r^2 \frac{\sin(kr)}{kr}\frac{\bar{\delta}_\mathrm{peak}(\tau_m, r)}{\bar{\delta}_\mathrm{peak}(\tau_m, 0)},
\end{aligned}
\label{eq:reconstruction_curvature_powerspectrum}
\end{equation}
where we choose the typical scale of the perturbation as the decoherence radius, i.e., $r_\mathrm{dec}=r_t$. In the following we concentrate on wavemodes ranging from~$k_t=r_t^{-1}$ to $5k_t$, which we expect to be the modes relevant for the collapse. Alternatively, we will also consider the ``almost scale-invariant'' power spectrum~$\mathcal{P}_\zeta(k)=k^3P_\zeta(k)/(2\pi^2)$, where~$P_\zeta$ is obtained from equation~\eqref{eq:reconstruction_curvature_powerspectrum}.

Regarding the window functions, according to \S~\ref{subsec:filtering_random_fields}, we will have a window function on super-horizon scales, implying that the second term on the RHS of equation~\eqref{eq:reconstructed_power_spectrum} becomes negligible, and one,~$\widehat{W'}_s$, on scales much smaller than the scales of the peak, which appear in the LHS of equation~\eqref{eq:reconstruction_curvature_powerspectrum}. In the range of wave modes of interest here~$\widehat{W'}_s(k)\equiv 1$. We approximate the non-linear transfer function~$\widehat{\mathcal{T}}_\mathrm{NL}$ as the linear transfer function corrected by the numerical factors found using numerical simulations and reported in \S~\ref{subsec:pbhs_abundance}.

The statistical properties of the field that generates the collapsing peaks can be  estimated given the assumed shape of the peaks. The spectral moments in equation~\eqref{eq:spectral_moments} can be computed at any time as
\begin{equation}
\begin{aligned}
\sigma^2_0(\tau) &= \left. \int \frac{d^3k}{(2\pi)^3} P_s(k,\tau) e^{-i \mathbf{k}\cdot\mathbf{r}} \right|_{\mathbf{r}=\mathbf{0}} = \xi_s(\tau,\mathbf{0}),\\
\sigma^2_1(\tau) = \cdots &= - \nabla^2 \xi_s(\tau,\mathbf{0}), \qquad \sigma^2_2(\tau) = \cdots = \nabla^2 \nabla^2 \xi_s(\tau,\mathbf{0}), 
\end{aligned}
\end{equation}
therefore, using equation~\eqref{eq:smoothed_correlation_function_delta_peaks}, we can write the spectral parameters in equation~\eqref{eq:spectral_parameters} as
\begin{equation}
\gamma(\tau) = - \frac{\nabla^2 \delta_\mathrm{peak}(\tau,\mathbf{0})}{\sqrt{\delta_\mathrm{peak}(\tau,\mathbf{0}) \nabla^2 \nabla^2 \delta_\mathrm{peak}(\tau,\mathbf{0})}}, \qquad R_\star(\tau) = \sqrt{-3\frac{\nabla^2 \delta_\mathrm{peak}(\tau,\mathbf{0})}{\nabla^2 \nabla^2 \delta_\mathrm{peak}(\tau,\mathbf{0})}}.
\end{equation}
Finally, for every peak profile, we can derive the variance of the overdensity field at horizon crossing~$\sigma_0(\tau_m)$ generating a given fraction of dark matter in PBHs by using equations~\eqref{eq:pbh_abundance} and~\eqref{eq:pbh_abundance_II}:
\begin{equation}
\begin{aligned}
f_\mathrm{PBH} \rho_{0c} \Omega_{\mathrm{dm},0} &= \frac{\mathcal{K}'(\alpha) M_\mathrm{hor}(\tau_m) \left[\sigma_0(\tau_m)\right]^{\upgamma_\mathrm{crit}}}{4 \pi^2 R^3_\star(\tau_m)} \times \\
&\qquad \times \int_{\nu_{c,m}^{(\alpha)}}^{\infty} d\nu \left(\nu -\nu_{c,m}^{(\alpha)}\right)^{\upgamma_\mathrm{crit}} G\left(\gamma(\tau_m),\gamma(\tau_m)\nu\right) e^{-\nu^2/2},
\end{aligned}
\label{eq:sigma0_equation}
\end{equation}
where the variance~$\sigma_0(\tau_m)$ appears also in the expression for the critical threshold~$\nu_{c,m}^{(\alpha)}$, computed using the peak height obtained in numerical simulations. Notice that the~$g_\star$ factors simplifies when combining equations~\eqref{eq:pbh_abundance} and~\eqref{eq:pbh_abundance_II}. The variance~$\sigma_0$ obtained from the equation~\eqref{eq:sigma0_equation} is consistently $10-30\%$ smaller than the typical Press-Schechter-like estimate~$\sigma^\mathrm{PS}_0 = \delta_{\mathrm{peak},0,c}/\left(\sqrt{2}\mathrm{Erfc}^{-1}(\beta_f)\right)$, where~$\mathrm{Erfc}^{-1}$ is the inverse of the complementary error function, for all the profiles and masses of interest. Therefore, using Press-Schechter overestimates the true amplitude of the curvature power spectrum by a factor~$20-70\%$ with respect to the prediction from Peak Theory. 

\begin{figure}[t]
\centerline{
\includegraphics[width=1.0\columnwidth]{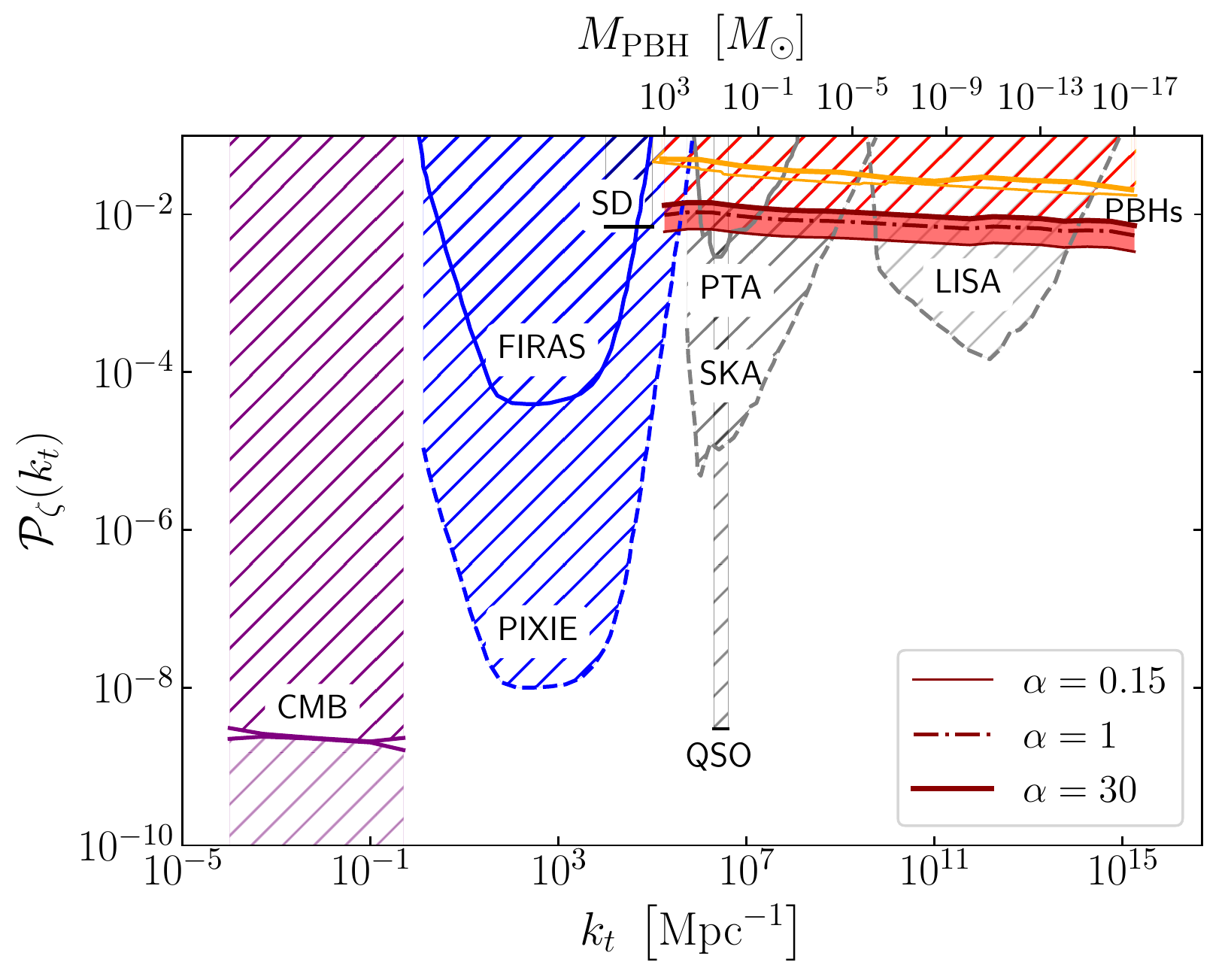}}
\caption{Maximum allowed  amplitude of the primordial curvature power spectrum. Current (solid lines) and forecasted (dashed lines) constraints from the cosmic microwave background (CMB, $1\sigma\ \mathrm{CL}$)~\cite{akrami:planckng2018}, spectral distortions (FIRAS and PIXIE)~\cite{chluba:powerspectrumconstraints}, gravitational waves (PTA, SKA and LISA)~\cite{byrnes:steepestgrowth}, Silk damping (SD)~\cite{jeong:powerspectrumconstraints}, quasar light curves (QSO)~\cite{karami:powerspectrumconstraints} and PBHs (the thin and thick orange lines correspond to Ref.~\cite{josan:powerspectrumconstraints} and \cite{satopolito:powerspectrumconstraints}, respectively). The red shaded region is the result of this work. It shows upper limits from PBH abundance for the range of profiles with shape parameter~$\alpha\in \left[0.15,30.0\right]$, considering the most recent constraints on the maximum allowed fraction of PBHs. We report our constraint for PBHs masses ranging from~$M_\mathrm{PBH}=10^{-17}\ M_\odot$ ($k_t\simeq 10^{15}\ \mathrm{Mpc}^{-1}$) to~$M_\mathrm{PBH}=10^{3}\ M_\odot$ ($k_t\simeq 10^{5}\ \mathrm{Mpc}^{-1}$).}
\label{fig:maximum_amplitude_powerspectrum}
\end{figure}

In this work we consider CPBHs with masses between~$10^{-17}\ M_\odot$ and~$10^3\ M_\odot$, even if our method applies also to different mass ranges. This mass range includes all the PBHs which have not evaporated by the present-day and for which we have observational constraints. We assume that all the CPBHs share the same formation time, hence that the primordial curvature power spectrum has a localised feature such as a spike. To connect typical scales and the compact object mass, we assume that all the CPBHs exceeded the critical threshold for formation by the same amount, which we choose to be~$(\delta_I-\delta_{I,c})=0.01$, generating CPBHs with masses~$M_\mathrm{PBH}=M_\mathrm{hor}(\tau_m)$, leaving the estimate of the CPBH initial mass function for future work. Following our conventions, the CPBHs mass is given by
\begin{equation}
M_\mathrm{PBH} = M_\mathrm{hor}(\tau_m) = \frac{a^2_\mathrm{eq}}{4t_\mathrm{eq}} k_t^{-2},
\label{eq:mass_scale_relation}
\end{equation}
where $a_\mathrm{eq}$ and $t_\mathrm{eq}$ are the scale factor and cosmic time at matter-radiation equality, respectively, and $k_t$ is related to the typical scale of the collapsing perturbation (see \S~\ref{subsec:filtering_random_fields}). Notice that relaxing this assumption does not have any impact on the constraint itself, in fact different choices of~$(\delta_I-\delta_{I,c})$ induce a rescaling in the relation linking the CPBH mass to the characteristic scale of the perturbation that generate it. Notice that a factor~$10$ of difference in~$(\delta_I-\delta_{I,c})$ generates a factor~$3$ of difference in~$M_\mathrm{PBH}$, therefore the connection between typical scales of the perturbation and the CPBHS masses is not extremely sensitive to changes in the value of the critical threshold.

In the following we will explicitly consider two extreme and one intermediate cases, $\alpha=0.15$, $\alpha=1$ and~$\alpha=30$, corresponding to very steep, a so-called ``Mexican-hat'' shape and very flat peaks, respectively.  In figure~\ref{fig:maximum_amplitude_powerspectrum}  we show the maximum amplitude of the primordial curvature power spectrum computed using equation~\eqref{eq:reconstruction_curvature_powerspectrum} along with the previous upper bound obtained using an approximated version of this procedure~\cite{josan:powerspectrumconstraints, satopolito:powerspectrumconstraints} and current and future upper bounds coming from different observables\footnote{To compute certain constraints, for instance those coming from spectral distortions and GWs, it is necessary to assume a shape of the primordial curvature power spectrum. We refer the interested reader to Ref.~\cite{byrnes:steepestgrowth}, where the change in the constraints assuming different power spectra shapes is shown. Since many of these constraints are forecast and the specifics of the instruments are unknown, in this work we use the curves found in Ref.~\cite{byrnes:steepestgrowth}, obtained assuming a curvature power spectrum that grows as~$\mathcal{P}_\zeta(k)\propto k^{4}$.}. This should be interpreted as the upper envelope of a family of spikes in the primordial curvature power spectrum each of which generates CPBHs of a given (monochromatic) mass falling in the range $10^{-17}\ M_\odot <M_\mathrm{PBH}< 10^{3}\ M_\odot$.

Compared to previous analyses, our more accurate procedure, which reduces the number of assumptions, gives stronger constraints on the maximum amplitude of the power spectrum. The improvement is approximately one order of magnitude on the scales of interest, from~$10^5\ \mathrm{Mpc}^{-1}$ to $10^{15}\ \mathrm{Mpc}^{-1}$, with tighter constraints for steeper profiles. In figure~\ref{fig:maximum_amplitude_powerspectrum} the red band includes all the peaks profiles considered here, for the maximum fraction of dark matter in PBHs~$f^\mathrm{max}_\mathrm{PBH}$ allowed by observations (see e.g., figure~1 of Ref.~\cite{bartolo:secondordergws}).

Even if the observational limits on~$f_\mathrm{PBH}(M_\mathrm{PBH})$ are very irregular and vary of several orders of magnitude between different masses, these differences are almost erased in figure~\ref{fig:maximum_amplitude_powerspectrum} because at leading order~$\sigma_0\propto \left(-\log f_\mathrm{PBH}\right)^{-1/2}$, as can be estimated using the Press-Schechter result. Therefore the improvements on the modelling are much more important than improvements on the observational constraints. Moreover, given that abundance constraints for PBHs with extended mass distributions are typically of the same order of magnitude of those for monochromatic ones~\cite{bellomo:emdconstraints, carr:extendedmassdistributionI, carr:extendedmassdistributionII}, the use of the former will not shift significantly our predictions.

Figure~\ref{fig:maximum_amplitude_powerspectrum} also shows that the range of power spectrum amplitudes needed to generate PBHs as (a component of) the dark matter, can be probed by future experiments, as SKA and LISA. This enables interesting synergies between these different experiments and probes. For example, in case of PBHs detection, of, say, $\sim 1\ M_{\odot}$ by LIGO, if these are to be CPBHs then SKA should see the signature of the corresponding stochastic background of gravitational waves generated by large curvature fluctuations. A non-detection of this signal on the other hand would indicate a different origin for PBHs, such as generation by topological defects.

\begin{figure}[t]
\centerline{
\includegraphics[width=1.0\columnwidth]{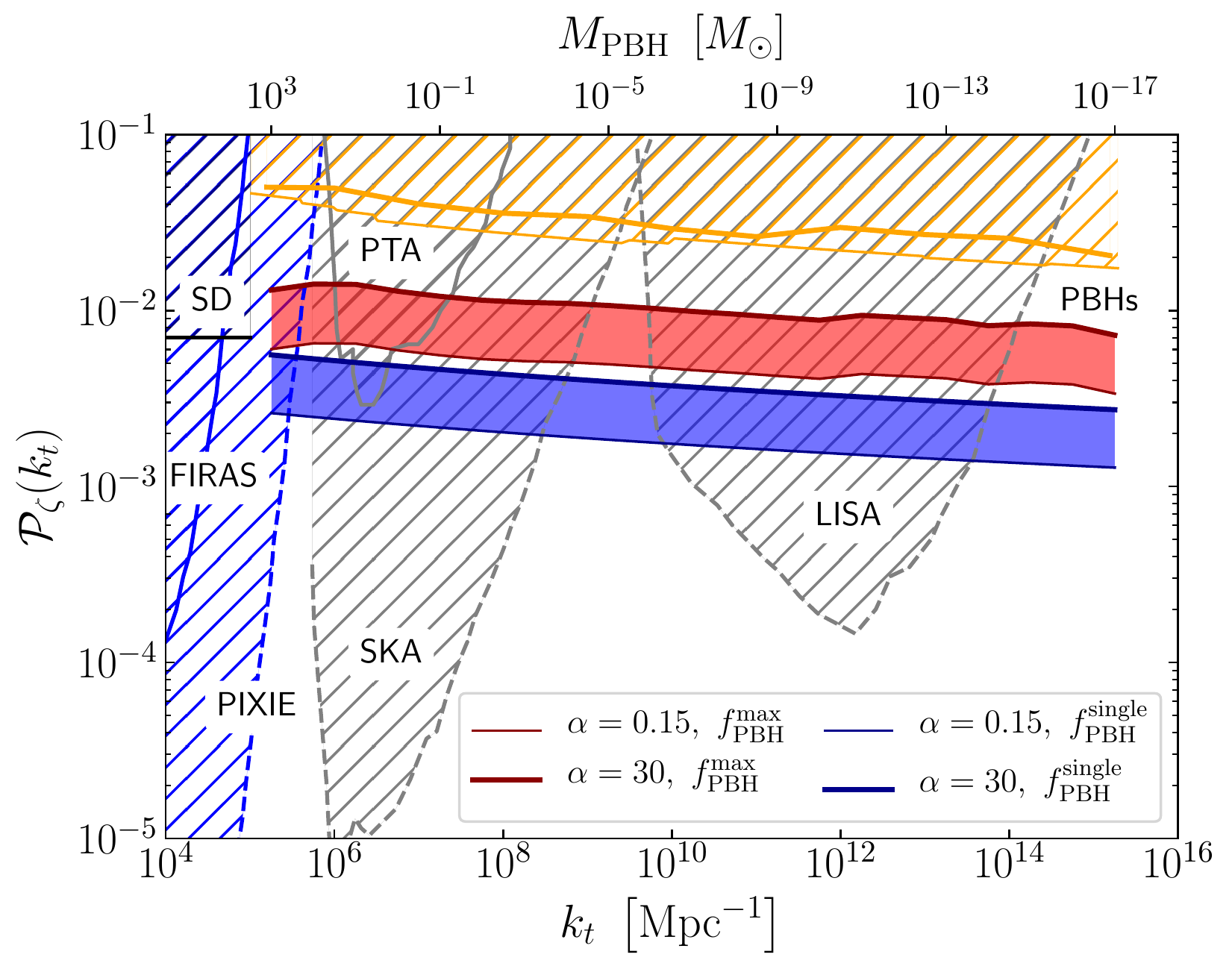}}
\caption{Same constraints of figure~\ref{fig:maximum_amplitude_powerspectrum}. Constraints from PBH abundance obtained using our methodology are indicated by the red shaded region and they assume the maximum abundance allowed by observations ($f_\mathrm{PBH}= f^\mathrm{max}_\mathrm{PBH}$), while the blue shaded region represents the constraints for the case where there is only one PBH in our Universe, i.e., $f_\mathrm{PBH}= f^\mathrm{single}_\mathrm{PBH}$.}
\label{fig:onepbh_constraint}
\end{figure}

On the other hand, the weak  sensitivity to the  abundance fraction~$f_\mathrm{PBH}$ already suggests that the existence of one single CPBH in our Universe is not compatible with a scale-invariant curvature power spectrum. The fraction of dark matter made of a single PBH can be written as
\begin{equation}
f^\mathrm{single}_\mathrm{PBH} = \frac{\rho_{\mathrm{PBH},0}}{\rho_\mathrm{dm,0}} = \frac{n_\mathrm{PBH} M_\mathrm{PBH}}{\rho_\mathrm{dm,0}} = \frac{M_\mathrm{PBH}/V_U}{\rho_\mathrm{dm,0}},
\label{eq:fPBH_min}
\end{equation}
where $V_U$ is the proper volume of the Universe\footnote{The proper volume of the Universe is given by $\displaystyle V_U=4\pi\int^{\infty}_0 dz \frac{\chi^2(z)}{(1+z)^3H(z)}\simeq 10^{11}\ \mathrm{Mpc}^3$, where~$\chi$ is the comoving distance, and it is approximately~$100$ times smaller than the comoving volume of the Universe $V_c = 4\pi\chi^3(\infty)/3 \simeq 10^{13}\ \mathrm{Mpc}^3$.}. Typical values of~$f^\mathrm{single}_\mathrm{PBH}$ ranges form~$f^\mathrm{single}_\mathrm{PBH}=3.2\times 10^{-39}$ for~$M_\mathrm{PBH}=10^{-17}\ M_\odot$ to~$f^\mathrm{single}_\mathrm{PBH}=3.2\times 10^{-19}$ for~$M_\mathrm{PBH}=10^{3}\ M_\odot$. In figure~\ref{fig:onepbh_constraint} we  show the minimum amplitude of the primordial power spectrum necessary for  generating a single PBH in the whole Universe, also in this case assuming a spike in the primordial curvature power spectrum.

Thus, even the existence of one single CPBH in the whole Universe is strongly incompatible, by orders of magnitude, with a simple scale invariant power spectrum at the level predicted by CMB observations (and extrapolated to these small scales). Therefore the detection of one single CPBH will reveal a completely different regime in the inflationary dynamic: it will indicate that the power spectrum has to rise from~$\mathcal{P}_\zeta\simeq 10^{-9}$ to~$\mathcal{P}_\zeta\simeq 10^{-3}-10^{-2}$, almost independently from the abundance of these objects, if they exist at all. Conversely, a null result by future experiments (SKA or LISA) in their target region of Fig.~\ref{fig:maximum_amplitude_powerspectrum} will rule out the possibility that PBHs might have formed via the collapse of primordial fluctuations.

\begin{figure}[t]
\centerline{
\includegraphics[width=1.0\columnwidth]{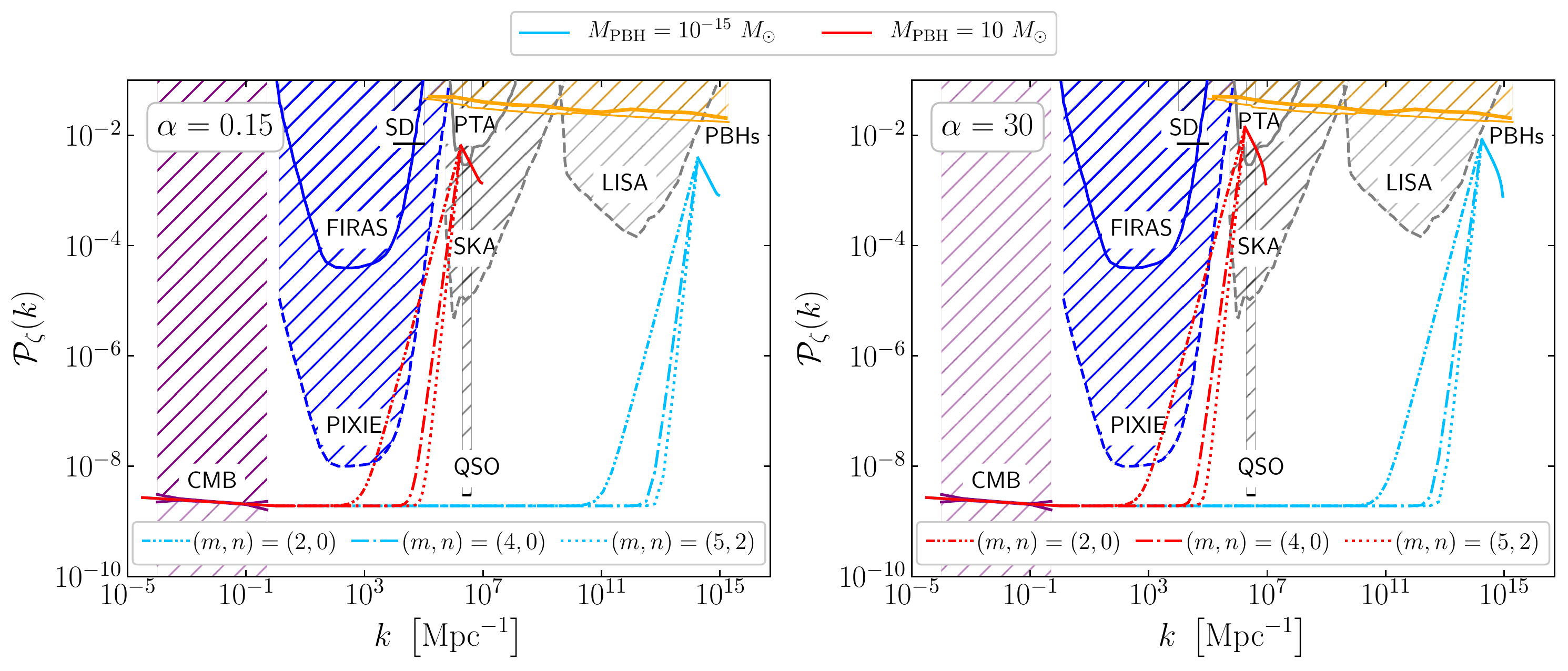}}
\caption{Shape of the reconstructed primordial curvature power spectra for steep (\textit{left panel}) and flat profiles (\textit{right panel}). In both cases we report the profile for~$M_\mathrm{PBH}=10^{-15}\ M_\odot$ (light blue line) and~$M_\mathrm{PBH}=10\ M_\odot$ (red line). To the right of the peak, the profile  is reconstructed using equation~\eqref{eq:reconstruction_curvature_powerspectrum}, and  to the left is computed assuming different model for the growth of the power spectrum (see text for details).}
\label{fig:shape_power_spectrum}
\end{figure} 

Moreover, by using equation~\eqref{eq:reconstruction_curvature_powerspectrum}, we can also compute the shape of the spike for modes~$k\gtrsim k_t$ comparable to or slightly larger than the typical mode~$k_t$. As we show in figure~\ref{fig:shape_power_spectrum}, the power spectrum to the right of the spike becomes increasingly steeper when~$\alpha$ increases, i.e., when the profile becomes flatter. Determining the shape of the power spectrum allow us to determine the shape parameter~$\alpha$, which together with the peak amplitude~$\mathcal{P}_\zeta(k_t)$ fix~$\sigma_0$, thus the abundance~$f_\mathrm{PBH}$ of CPBHs produced by the spike in the primordial curvature power spectrum. Therefore, without knowing the shape of the spike, it is not possible to uniquely determine if CPBHs form a relevant fraction of dark matter.

In figure~\ref{fig:shape_power_spectrum} we show possible shapes of the curvature power spectrum profile from cosmological scales up to the scale of the peak. At cosmological scales $(k\lesssim 1\ \mathrm{Mpc}^{-1})$ the primordial curvature power spectrum is very well constrained to be almost scale invariant, namely~$\mathcal{P}_\zeta(k) = A_s\left(k/k_\mathrm{pivot}\right)^{n_s-1}$, where $A_s$ is the scalar perturbations amplitude, $n_s$ is the scalar tilt and $k_\mathrm{pivot}$ is the pivot scale\footnote{According to the latest Planck collaboration results~\cite{aghanim:planckcosmoparams2018}, we have~$\log(10^{10}A_s) = 3.047 \pm 0.014$ and $n_s=0.9665 \pm 0.0038$, measured with $k_\mathrm{pivot}=0.05\ \mathrm{Mpc}^{-1}$.}. At intermediate scales~$(1\ \mathrm{Mpc}^{-1} \lesssim k \lesssim k_t)$, the primordial curvature power spectrum determines the clustering properties of CPBHs~\cite{chisholm:pbhclusteringI, chisholm:pbhclusteringII, alihaimoud:pbhclustering, desjacques:pbhclustering, ballesteros:pbhclustering, moradinezhad:pbhclustering, garriga:pbhclustering}. Since a full modelling of this  goes beyond the scope of this paper, we phenomenologically parametrise the power spectrum in this range of scales using the formula~$\mathcal{P}^{(m,n)}_\zeta(k)=B k^m \log^n (k) + C$, where~$B$ and~$C$ are fitting parameters. Ref.~\cite{byrnes:steepestgrowth} claimed that the choice~$(m,n)=(4,0)$ represents the steepest rise of the primordial power spectrum in the context of single-field inflation, however it was later shown that an even steeper rise, characterized by the parameters~$(m,n)=(5,2)$, is also possible~\cite{carrilho:steepestgrowth}. We show in figure~\ref{fig:shape_power_spectrum} both options, along with a third one, characterized by a milder rise~$(m,n)=(2,0)$. These different curves effectively change the clustering properties of CPBHs which can in principle be determined observationally. As stated previously, to compute some of these constraints we need to assume a shape of the curvature power spectrum. The constraints reported in the figure are obtained in Ref.~\cite{byrnes:steepestgrowth} for the~$(4,0)$ case. While we expect that in the~$(5,2)$ case the constraints do not change significantly, the amount of change in the shallower growth case~$(2,0)$ might be larger.

In conclusion, we summarize here the relevant steps to follow to reconstruct both amplitude and shape of the primordial curvature power spectrum:
\begin{enumerate}
\item choose an overdensity peak profile. In principle this should not be an arbitrary choice but in the absence of a complete prediction for the peak profile we advocate using one from the family of equation~\eqref{eq:K_profile};
\item use peak theory's result in equation~\eqref{eq:smoothed_correlation_function_delta_peaks} to connect the overdensity peak profile and the two-point correlation function, thus the power spectrum combining equations~\eqref{eq:nongaussian_smoothed_power_spectrum} and~\eqref{eq:reconstructed_power_spectrum};
\item in absence of a exact prediction of the non-linear transfer function~$\widehat{\mathcal{T}}_\mathrm{NL}$, correct the linear one using the numerical factors calibrated on the numerical simulations. For the family of profiles we adopted, these values reads as~$\widehat{\mathcal{T}}_\mathrm{NL}/\widehat{\mathcal{T}}_\mathrm{LIN} \simeq 1.8,\ 1.5,\ 1.2$ for profiles characterised by~$\alpha=0.15,\ 1.0,\ 30.0$;
\item after estimating the spectral parameters~$\gamma(\tau_m)$ and~$R_\star(\tau_m)$ from the profile shape, compute the variance at horizon crossing~$\sigma_0(\tau_m)$ by solving equation~\eqref{eq:sigma0_equation}. The amplitude of the peak (in the $\nu$ variable) has to be computed using numerical simulations. The PBHs abundance value $f_{\rm PBH}$ has to be set for the mass of the corresponding  compact object which is related to the  horizon crossing time  via, for instance, equation~\eqref{eq:mass_scale_relation};
\item finally, the peak maximum amplitude of figures~\ref{fig:maximum_amplitude_powerspectrum} and~\ref{fig:onepbh_constraint} is obtained evaluating equation~\eqref{eq:reconstruction_curvature_powerspectrum} at the wavemode~$k_t=r_t^{-1}$;
\item
the peak shape (e.g., red and cyan lines in figure~\ref{fig:shape_power_spectrum}) is obtained by evaluating equation~\eqref{eq:reconstruction_curvature_powerspectrum} at wavemodes~$k_t\leq k \leq 5k_t$ for scales smaller than the typical scale of the perturbation.  For scales greater than the typical scale of the perturbation, match the  $\mathcal{P}_\zeta$ at maximum to one of our~$\mathcal{P}^{(m,n)}_\zeta$ models.
\end{enumerate}


\section{Conclusions}
\label{sec:conclusions}
In the model where Primordial Black Holes (PBHs) form from large primordial curvature perturbations, CPBHs, PBH abundance can be used to set limits on the amplitude of the primordial power spectrum of perturbations on scales that are not easily accessible by other probes. However, making this connection requires a detailed modelling of PBHs collapse and formation in a cosmological context. We improve upon previous literature by eliminating a series of approximations used so far. It turns out that the accuracy in the modelling is (much) more important than the precision on the constraints on the PBH abundance, further motivating our effort.

In this work we set the connection between primordial power spectrum and PBH abundance on solid theoretical grounds. To achieve this goal we have, for the first time, combined three key inputs to the problem: \textit{(i)} the numerical relativistic simulations, to assess the conditions under which CPBHs of a given mass form; \textit{(ii)} the cosmology connection, to link the properties of individual overdensity peaks able to create CPBHs to the statistics of the underlying cosmological random field; and \textit{(iii)} Peak theory, to interpret PBH abundance in terms of a primordial amplitude of the power spectrum of a cosmological density field.

Our major results can be summarised as follows. The first four results are methodological, the last two are new constraints.
\begin{enumerate}
\item  Full non-linear results for the evolution of a curvature perturbation must be used, for which numerical simulations are crucial. Even if at initial time the $\zeta$-curvature perturbation (eventually giving rise to a CPBH) and its gradients are small (and therefore the equations can be linearised), at horizon crossing this is not the case any more. In fact, using a linear approximation underestimates the real size of the perturbation (and hence the mass enclosed in the horizon) by a relatively large factor (up to~$\sim 6$), depending on the shape of the perturbation: steeper perturbation profiles are most dramatically affected (see~\S~\ref{sec:pbh_formation}).
 
\item There are three scales involved in the problem to be compared to the horizon crossing scale of the perturbation. Two of which are physical, one is a mathematical requirement. One (small) scale is the one necessary to define a smoothing scale to make the underlying overdensity field at least differentiable and to define in it peaks and troughs. This scale is also necessary to define finite spectral moments of the field. One larger scale is the size of the sound horizon, below which pressure gradients smooth out perturbations. Finally scales well above the horizon at all times during the collapse should not influence the dynamics. The first scale is not physical, it is a mathematical operation and we have some freedom to decide what ``smoothing'' should be used. We argue that it should be smaller than the typical scale of the perturbation of interest in such a way that its specific choice should not influence the description of the dynamical evolution of a perturbation. The second scale is instead physical: pressure effects damp perturbations on scales smaller than the sound horizon. At horizon crossing this is comparable to the size of the perturbation itself. In our approach this is accounted for as the evolution of the collapse is modelled numerically (see appendix~\ref{app:numerical_simulations} for details). Finally, scales that are well above the horizon at all times during formation do not influence sub-horizon dynamics. The collapse is not instantaneous so perturbations of scales above but comparable to the horizon at a given time might influence the collapse at a later time. Nevertheless, ultra-long scale perturbations associated to modes well above the horizon throughout the CPBH formation and collapse must appear as a constant background, hence should not play any role in determining whether a CPBH forms or not. 

\item Non-linearities are important also in the process of smoothing. In this work we recommend to perform the smoothing in physical coordinates, to correctly include the fact the curvature might be not negligible. In this sense, there is an ambiguity on which field should be smoothed, since smoothing the curvature or the overdensity field is not equivalent, as it is in linear theory. Moreover, the importance of non-linearities suggests that also non-Gaussianities might be important, in fact we proved in \S~\ref{subsec:nongaussianity} that the two-point function of the overdensity field receives contributions from all the $n$-point functions of the $\zeta$-curvature field because of the non-linear relation between overdensity and curvature.

\item While numerical simulations can follow one perturbation at the time, the Universe is filled by a superposition of perturbations. Peak theory connects the statistics of a smoothed (Gaussian) random field defined by its power spectrum to the statistical distribution of its local maxima (above a given threshold). By identifying these local maxima with peaks of initial overdensity perturbations, the results from numerical simulations (especially the conditions on the peak height for collapse to a BH) can be used to derive the abundance of collapsed objects. Not all the peaks correspond to sites where a CPBH form; numerical simulations are key in defining a threshold criterion to assess which peaks collapse and which do not. This criterion depends also on the shape of the initial perturbation. Peak theory then enables us to connect the average shape of the peaks and its variance to the statistical properties of the random field and thus make a statistical connection to the numerical simulation results. Future improvements in Peak theory, e.g., accounting for the ``black-hole-in-black-hole'' problem for models characterised by a very flat power spectrum, will certainly provide an even more accurate estimate of CPBHs abundance. Inclusion of non-Gaussianities will also contribute to increase the accuracy.

\item We show in figure~\ref{fig:maximum_amplitude_powerspectrum} that the modelling done in the previous steps is fundamental in determining the correct constraint on the maximum amplitude of the primordial curvature perturbations power spectrum. In particular, our more accurate approach, which resorts to less approximations, for instance in the estimation of the variance or of the window and transfer functions, yields constraints are one order of magnitude tighter than what was previously estimated, for the entire range of modes or, equivalently, for a wide range of CPBHs masses.

\item The existence of CPBHs generated from primordial perturbations, we confirm, is incompatible with a scale-invariant power spectrum as measured at cosmological scales (see figure~\ref{fig:onepbh_constraint}). Moreover we show that the detection of one single CPBHs would signal a significant departure from the standard slow-roll inflationary scenario.

\item The method presented in this work provides also an alternative way to probe the formation mechanism of PBHs. If PBHs are detected and no boost in the primordial curvature power spectrum is found by SKA or LISA, for instance by detecting a gravitational waves background generated by the same large scalar perturbation that created the PBHs, then an alternative PBH formation mechanism must be at play, e.g., cosmic topological defects. In this context it is crucial to find new ways to probe the origin (end-point of stellar evolution or primordial) of BHs detected by current and future gravitational waves observatories, for instance cross-correlating galaxy and gravitational waves maps~\cite{raccanelli:pbhprogenitors, raccanelli:gwastronomy, scelfo:gwxlss}, measuring BHs binaries eccentricity~\cite{cholis:orbitaleccentricities}, the BH mass function~\cite{kovetz:pbhmassfunction, kovetz:pbhandgw} and the BHs merger rate~\cite{nakamura:pbhmergerrate, alihaimoud:pbhmergerrate}, or using fast radio burst~\cite{munoz:fastradioburst}.

\item While here we have concentrated on scales comparable to the typical peak size, in principle our method can be extended to constrain larger scales, in the intermediate regime between standard cosmological scales and typical peaks size, via primordial clustering of PBHs~\cite{chisholm:pbhclusteringI, chisholm:pbhclusteringII, alihaimoud:pbhclustering, desjacques:pbhclustering, ballesteros:pbhclustering, moradinezhad:pbhclustering, garriga:pbhclustering}. We have illustrated this in figure~\ref{fig:shape_power_spectrum} and will be presented elsewhere.
\end{enumerate}

This paper highlights that the details of the connection between the limits on the PBHs abundance,~$f_\mathrm{PBH}$, and the primordial curvature power spectrum are much more important than the limits on the abundance themselves. Nevertheless many of the results presented in this work can be applied to estimate the CPBHs abundance from a given primordial curvature power spectrum, i.e., the way back. In particular, our work covers some key aspects up to the time of CPBHs formation. However there are many others subtleties involved in that estimation, for instance the modelling of processes involving the CPBHs from the time of formation to today, e.g., the modelling of CPBHs accretion or CPBHs clustering, that are not addressed. For this reason the way back is a very delicate issue. Even if our work does not treat those aspects, we believe it is of value as it still provides the first key steps to obtain the correct CPBHs abundance today.

To conclude, the results presented in this work represent a remarkable example of how both the existence and the non-existence of one of the most popular dark matter candidates can be used in cosmology. In particular, PBHs have the potential to probe~$10$ order of magnitude in terms of scales or, alternatively,~$20$ extra e-folds, shedding new light on the inflationary paradigm~\cite{bellomo:powerspectrum}.


\section*{Acknowledgments}
We thank Yashar Akrami, Christian Byrnes, Bernard Carr, Philippa Sarah Cole, Cristiano Germani, Anne Green and Sam Young for useful comments on the draft. Funding for this work was partially provided by the Spanish MINECO under projects AYA2014-58747-P AEI/FEDER, UE, and MDM-2014-0369 of ICCUB (Unidad de Excelencia Mar\'ia de Maeztu). AK was supported by Erasmus+Trainership scholaship and by MDM-2014-0369 of ICCUB (Unidad de Excelencia Mar\'ia de Maeztu). AK also acknowledges the Netherlands Organisation for Scientific Research (NWO) for support in the latest stages of the work. NBe is supported by the Spanish MINECO under grant BES-2015-073372. AR has received funding from the People Programme (Marie Curie Actions) of the European Union H2020 Programme under REA grant agreement number 706896 (COSMOFLAGS). NBa, DB and SM acknowledge partial financial support by ASI Grant No. 2016-24-H.0. IM has been supported by FPA2016-76005-C2-2-P,MDM-2014-0369 of ICCUB (Unidad de Excelencia Maria deMaeztu), AGAUR 2014-SGR-1474. LV acknowledges support by European Union's Horizon 2020 research and innovation programme ERC (BePreSySe, grant agreement 725327).


\appendix
\section{Numerical simulations}
\label{app:numerical_simulations}
The results for the threshold of PBH formation used in this paper to reconstruct the shape of the power spectrum, which allows PBHs to account for the whole dark matter, have been obtained with numerical simulation of gravitational collapse, stating from the initial conditions described in \S~\ref{subsec:super_horizon_regime}. The numerical code used is the same of Refs.~\cite{musco:pbhformation, polnarev:curvatureprofiles, musco:criticalcollapse, musco:selfsimilarity, musco:pbhthreshold}, which has been fully described previously and therefore we give only a brief outline of it here. 

It is an explicit Lagrangian hydrodynamics code with the grid designed for calculations in an expanding cosmological background. The basic grid uses logarithmic spacing in a mass-type comoving coordinate, allowing it to reach out to very large radii while giving finer resolution at small radii necessary to have a good resolution of the initial perturbation. The initial conditions -- initial data obtained as numerical solutions -- are specified on a space-like slice at constant initial cosmic time $t_\mathrm{ini}$ defined as $a_\mathrm{ini}H_\mathrm{ini}\tilde{r}_m = 10$ while the outer edge of the grid has been placed at $90 R_m$ (where $\tilde{r}_m$ and $R_m$ have been defined in \S~\ref{subsec:formation_criterion}), to ensure that there is no causal contact between it and the perturbed region during the time of the calculations. The initial data are evolved using the Misner-Sharp-Hernandez equations so as to generate a second set of initial data on an initial null slice which 
are then evolved using the Hernandez-Misner equations. During the evolution, the grid is modified with an adaptive mesh refinement scheme (AMR), built on top of the initial logarithmic grid, to provide sufficient resolution to follow black hole formation down to extremely small values of $(\delta_I-\delta_{I,c})$.

\begin{figure}[t]
\vspace{-1.5cm}
\centerline{
\includegraphics[width=0.53\textwidth]{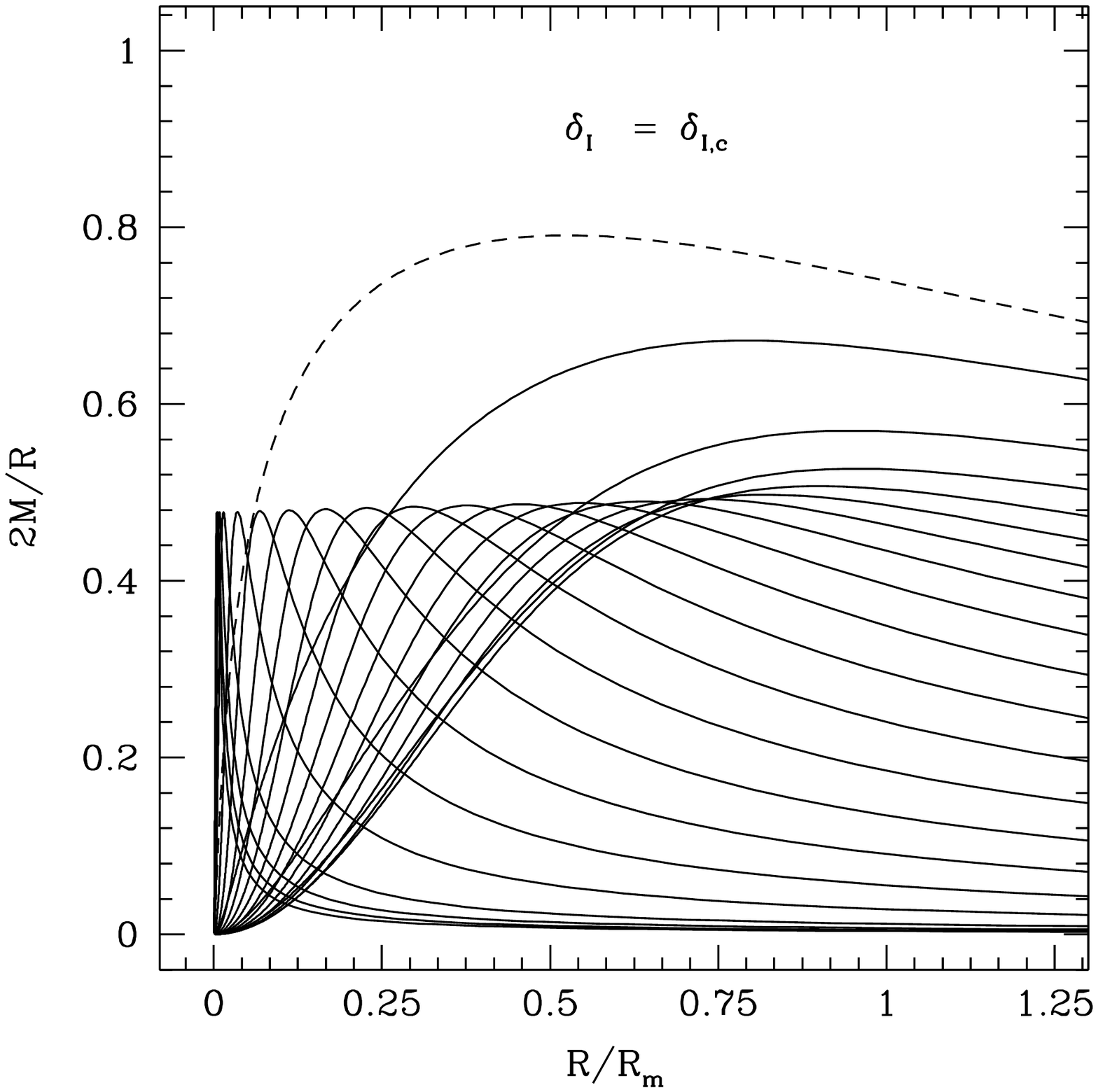} 
\includegraphics[width=0.53\textwidth]{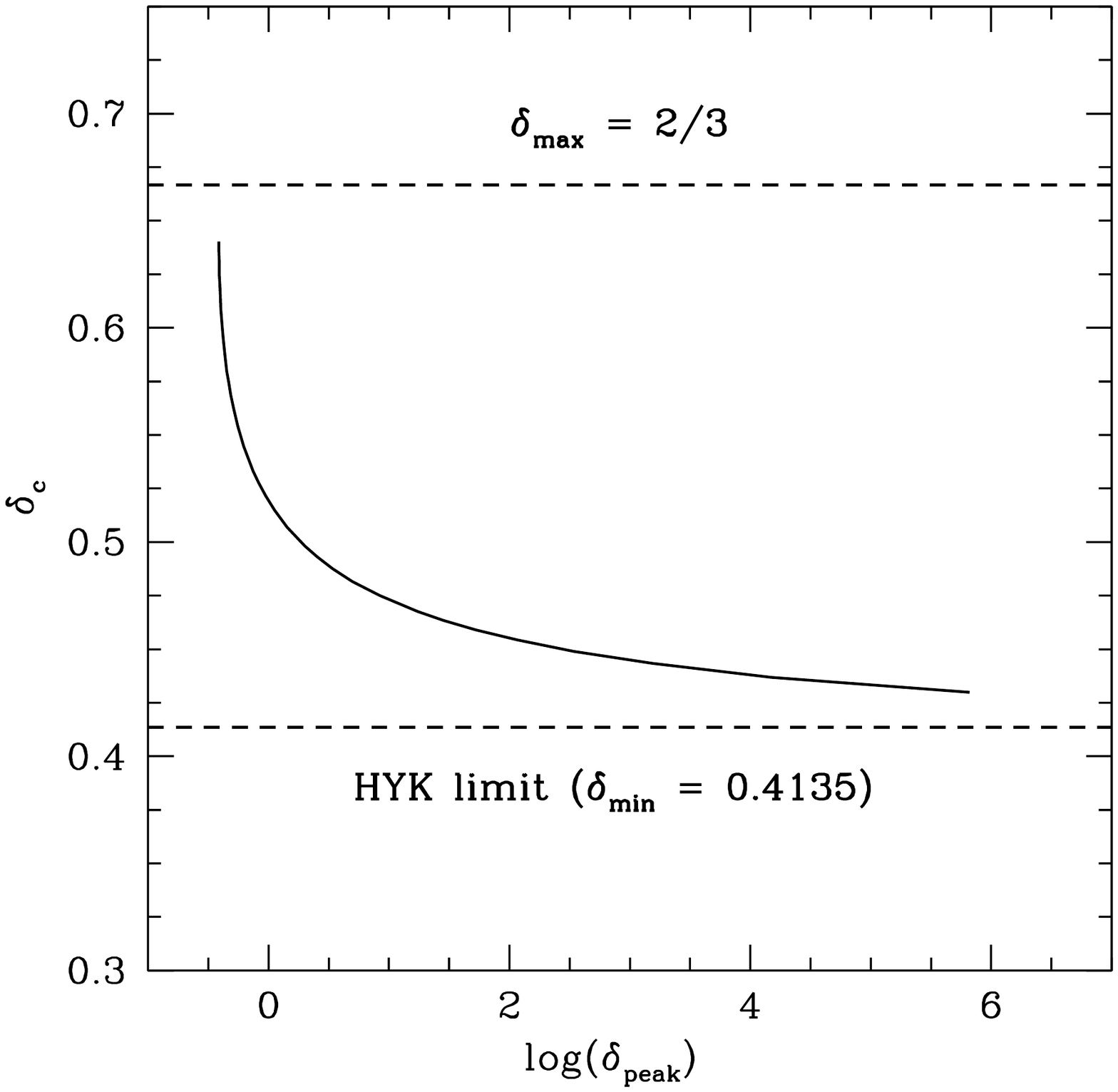}}
\vspace{-2.0cm}
\caption{\textit{Left panel:} dynamical behaviour of $2M/R$ against $R/R_m$ plotted at different time-slice for the critical solution of a zero mass black hole when $\delta_I=\delta_{I,c}$ obtained from equation~\eqref{eq:mass_critical_collapse} when~$\alpha=1$ (Mexican-Hat shape). The dashed line  corresponds to the initial time-slice, and the peak of $2M/R$ is initially decreasing when the perturbation is still expanding, reaching afterwards nearly equilibrium state moving inward when the perturbation is collapsing. Figure taken from Ref.~\cite{musco:criticalcollapse}. \textit{Right panel:} numerical behaviour of the critical threshold~$\delta_{I,c}$ against the corresponding behaviour of the critical peak amplitude~$\delta_{\mathrm{peak},0}$ for different shapes~$(0.15 \leq \alpha \leq 30)$. Figure taken from Ref.~\cite{musco:pbhthreshold}.}
\label{fig:app_critical}
\end{figure}

The critical threshold $\delta_{I,c}$ is found from the evolution of $2M/R$ as function of time, looking at the evolution of the peak of this ratio: when $\delta_I>\delta_{I,c}$, the peak is increasing during the collapse, reaching the condition for the apparent horizon $R(r,t)=2M(r,t)$ identifying the formation of a BH (see e.g., Ref.~\cite{helou:apparenthorizon, faraoni:foliation}), while when $\delta_I<\delta_{I,c}$, the peak is decreasing, no apparent horizon forms and the collapsing overdensity bounces into the 
expanding Friedmann-Robertson-Walker Universe. In the left panel of figure~\ref{fig:app_critical} we show the behaviour of $2M/R$ when $\delta_I\simeq\delta_{I,c}$, where the dashed line is the time slice of the initial conditions. During the first stage of the evolution the perturbation is still expanding and the peak is decreasing, as can be seen from the following time slices, while 
when the perturbation starts to collapse, the peak of $2M/R$ is in equilibrium, moving towards the centre with an almost constant value, because of the very close equilibrium between gravity and pressure one has at the threshold ($\delta_I\simeq\delta_{I,c}$). During this nearly equilibrium phase, matter is spread outward from a relativistic wind, keeping the shrinking region with an almost constant compactness (see Ref.~\cite{musco:criticalcollapse} for more details). 

As we have seen in \S~\ref{subsec:formation_criterion},~$\delta_I(r_m, t_m) \simeq (2M/R)_\mathrm{peak}$ and one can calculate the amplitude of $\delta_{I,c}$ looking for this equilibrium solution characterized by self-similar behaviour. This explains also the nature of the critical collapse characterizing PBH formation when~$\delta_I\simeq\delta_{I,c}$, with the mass spectrum given by the scaling law of equation~\eqref{eq:mass_critical_collapse} (see Ref.~\cite{musco:selfsimilarity} for more details).

In the right panel of figure~\ref{fig:app_critical} we can see the numerical relation between the threshold~$\delta_{I,c}$ and the corresponding critical value of the peak amplitude~$\delta_\mathrm{peak}$ for cosmological perturbations with the energy density profile obtained from equation~\eqref{eq:K_profile}, for the profile steepness parameter varying between~$\alpha=0.15$ (high~$\delta_{\mathrm{peak},0}$, low~$\delta_{I,c}$), and $\alpha=30$ (low $\delta_{\mathrm{peak},0}$, high $\delta_{I,c}$) corresponding to the lowest and the largest value of~$\alpha$ for the simulated profiles, respectively.

Because we are considering here only shapes characterized by one parameter, each value of~$\delta_{I,c}$ is associated to a different value of the peak amplitude~$\delta_{\mathrm{peak},0}$. The inverse behaviour between the two quantities is a key feature of the effect of the pressure gradients on the collapse, measuring the steepness of the profile, which can be simply measured by the ratio~$(\tilde{r}_0/\tilde{r}_m)=(3/2)^{1/2\alpha}$ (if the profile is characterized by more than one parameter the measure of the steepness is more complicated). In general, when the profile of the compaction function is steeper, which corresponds to a broad profile of the density contrast ($\alpha\gg 1$), as one can see from figures~\ref{fig:curvature_profiles} and~\ref{fig:critical_collapse}, the pressure gradients modify significantly the shape during the non-linear evolution of the collapse after horizon crossing. This gives a larger value of the threshold $\delta_{I,c}$ which accounts for the additional excess of mass necessary to compensate the mass that will be lost during the collapse, up to the maximum value of~$\delta_{I,c}=2/3$ corresponding to a top-hat shape ($\tilde{r}_0/\tilde{r}_m=1$). On the contrary, if the profile of the compaction function is not very steep ($\alpha\lesssim1$),  
which corresponds to a steep profile of the density contrast, the pressure gradients do not modify substantially the shape during the collapse, with a smaller value of~$\delta_{I,c}$, bounded by the numerical value $\delta_{I,c}\simeq 0.4135$ (Harada-Yoo-Kohri limit) found analytically in Ref.~\cite{harada:haradalimit}, where the role of the pressure gradients was neglected. The connection between the shape and the value of the threshold has been carefully analysed in Ref.~\cite{musco:pbhthreshold}.


\section{Counting the relativistic degrees of freedom}
\label{app:degrees_of_freedom}
In this appendix we report the exact calculation of how the energy density of radiation scales from the Early Universe until today, since it is often presented in an approximated version. We review the main steps to derive equation~\eqref{eq:pbh_abundance_II}, more details can be found in several classical books, see e.g., Ref.~\cite{kolb:dofs}.

Consider a thermal bath of photons with temperature~$T_\gamma$. Deep in radiation-dominated era there were other relativistic species in thermal equilibrium with photons (at least all or part of Standard Model particles, depending on the temperature), each characterised by~$g_j$ internal degrees of freedom. In principle the existence of other relativistic particles decoupled from photons is possible, hence in the following we account also for them, assuming that they have a thermal distribution with temperature~$T_j\neq T_\gamma$. The energy density and entropy density of the entire fluid read as
\begin{equation}
\rho_\mathrm{rad}(T_\gamma) = \frac{\pi^2}{30} \frac{(k_B T_\gamma)^4}{(\hbar c)^3} g_{\star, \rho}(T_\gamma), \qquad s_\mathrm{rad}(T_\gamma) = \frac{2\pi^2}{45} \frac{k^4_B T_\gamma^3}{(\hbar c)^3} {g_{\star, s}(T_\gamma)},
\label{eq:app_energydensity_entropydensity}
\end{equation}
where~$k_B$ is the Boltzmann constant, $\hbar$ is the reduced Planck constant, $c$ is the speed of light and the total number of effective degrees of freedom for energy and entropy densities are defined by 
\begin{equation}
\begin{aligned}
g_{\star, \rho}(T_\gamma) &= \sum_{\mathrm{bosons}} g_j \left(\frac{T_j}{T_\gamma}\right)^4 + \frac{7}{8} \sum_{\mathrm{fermions}} g_j \left(\frac{T_j}{T_\gamma}\right)^4, \\
g_{\star, s}(T_\gamma) &= \sum_{\mathrm{bosons}} g_j \left(\frac{T_j}{T_\gamma}\right)^3 + \frac{7}{8} \sum_{\mathrm{fermions}} g_j \left(\frac{T_j}{T_\gamma}\right)^3,
\label{eq:app_degrees_of_freedom}
\end{aligned}
\end{equation}
where the sum runs over relativistic species only because their contribution dominates over that of non-relativistic ones.

Using the conservation of entropy~$g_{\star, s} a^3 T^3 = {const.}$ we find the scaling of temperature in an expanding Universe. Therefore the energy density of radiation at any time can be consistently related to the radiation energy density today~$\rho_\mathrm{rad}(T_0)$ writing equation~\eqref{eq:app_energydensity_entropydensity} as
\begin{equation}
\rho_\mathrm{rad}(T_\gamma) = \rho_\mathrm{rad}(T_0) \left(\frac{T_\gamma}{T_0}\right)^4 \frac{g_{\star, \rho}(T_\gamma)}{g_{\star, \rho}(T_0)} = \rho_\mathrm{rad}(T_0) \frac{g_{\star, \rho}(T_\gamma)}{g_{\star, \rho}(T_0)} \left(\frac{g_{\star, s}(T_0)}{g_{\star, s}(T_\gamma)}\right)^{4/3} \left(\frac{a_0}{a_\gamma}\right)^4,
\label{eq:app_energydensity_scaling}
\end{equation}
where~$T_0$ is the photon temperature today, $a_0$ and~$a_\gamma$ are the scale factors today and of when photons had temperature~$T_\gamma$, respectively. This relation is then used to obtain equation~\eqref{eq:pbh_abundance_II}. Notice that the approximation~$g_{\star, \rho}\sim g_{\star, s}$ has often been taken in the literature.

Now we want to consider the neutrino contribution to equations~\eqref{eq:app_energydensity_scaling} and~\eqref{eq:pbh_abundance_II}. Even if we have not measured neutrino masses yet, we know from neutrino oscillation that at least two of them are massive and we measured the mass gap between different mass eigenstates~\cite{gonzalez:masssplitting}:
\begin{equation}
m_2^2-m_1^2 = 75\ (\mathrm{meV})^2, \qquad |m_3^2 - m_l^2| = 2519\ (\mathrm{meV})^2,
\end{equation}
where $m_l=m_1$ in the normal hierarchy scenario~$(m_1<m_2<m_3)$ while~$m_l=m_2$ in the inverted hierarchy scenario~$(m_3<m_1<m_2)$.  Massive neutrinos become non-relativistic around redshift~$1+z_\mathrm{nr}\simeq 2 \times \left[m_\nu/(1\ \mathrm{meV})\right]$~\cite{lesgourgues:neutrinos}, therefore in the past at least two neutrinos became non-relativistic, even when the lightest mass eigenstate is massless, i.e., when $m_1=0$ and~$m_3=0$ for normal and inverted hierarchies, respectively. 

In the following we assume that all the neutrinos became non-relativistic, since this happens even for reasonably low values of the lightest state, e.g., $m_\nu\simeq 1\ \mathrm{meV}$. Therefore, when considering the energy density in radiation today, we have to include only photons, corresponding to~$g_{\star, \rho}(T_0)=2$. 

On the other hand, when estimating the degrees of freedom for the entropy we have to be more careful, in fact the entropy conservation argument $g_{\star, s}(T_\gamma) a^3_\gamma T_\gamma^3 = g_{\star, s}(T_\mathrm{nr}) a_\mathrm{nr}^3 T_\mathrm{nr}^3$ can be used until when neutrinos were relativistic, the temperature of the photon bath was~$T_\mathrm{nr}$ and the scale factor~$a_\mathrm{nr}$. Afterwards they will not contribute to the entropy, however they ``disappear'' without warming the photons, as it happens with particles annihilation. Therefore after the non-relativistic transition of neutrinos, photon temperature evolves as~$a_\mathrm{nr}^3 T_\mathrm{nr}^3 = a_0^3 T_0^3$. For this reason, it is more accurate to report~$g_{\star, s}(T_\mathrm{nr})$ in equation~\eqref{eq:app_energydensity_scaling} or, alternatively, to compute~$g_{\star, s}(T_0)$ considering neutrino as relativistic particles, i.e., $g_{\star, s}(T_0)=3.909$. 

\begin{table}[t]
\centerline{
\begin{tabular}{|c|c|c|}
\hline
\begin{tabular}{@{}c@{}} Number of Relativistic \\ Neutrinos Today \end{tabular} 	&	$g_{\star, \rho}(T_0)$	&	$g^{4/3}_{\star, s}(T_0)$	\\
\hline
\hline
$0$	&	$2.00$	&	$2.52$	\\
$1$	&	$2.45$	&	$3.64$	\\
$2$	&	$2.91$	&	$4.86$	\\
$3$	&	$3.36$	&	$6.16$	\\
\hline
\end{tabular}}
\caption{Total number of effective degrees of freedom for energy density~$g_{\star, \rho}(T_0)$ and entropy density~$g^{4/3}_{\star, s}(T_0)$	 today. Values are obtained using equation~\eqref{eq:app_degrees_of_freedom}, neutrino temperature~$T_\nu/T_\gamma=(4/11)^{1/3}$ and~$g_\nu=2$ for every neutrino family.}
\label{tab:degrees_of_freedom}
\end{table}

We report in table~\ref{tab:degrees_of_freedom} the total number of effective degrees of freedom for energy and entropy density. As it can be seen, the relative difference in assuming~$0$ or~$1$ relativistic neutrinos today is~$23\%$ and~$44\%$ for~$g_{\star, \rho}$ and~$g^{4/3}_{\star, s}$, respectively. On the other hand, the relative difference we have assuming~$3$ relativistic neutrino becomes~$68\%$ and~$144\%$ for~$g_{\star, \rho}$ and~$g^{4/3}_{\star, s}$, respectively, compared to the case of no relativistic neutrinos today. In our calculation we use $g_{\star, s}(T_0)=3.909$ and $g_{\star, \rho}(T_0)=2.0$.


\bibliography{biblio}
\bibliographystyle{utcaps}


\end{document}